\documentclass[12pt,onecolumn]{IEEEtran}
\usepackage{epsf}
\usepackage{graphicx}% Extended graphics
\usepackage{amsmath} \usepackage{amssymb}
\usepackage{booktabs}
\allowdisplaybreaks

\newtheorem{theorem}{Theorem}[section]

\newtheorem{remark}{Remark}[section] 

\long\def\symbolfootnote[#1]#2{\begingroup%
\def\thefootnote{\fnsymbol{footnote}}\footnote[#1]{#2}\endgroup}

 \def\DD{{\cal D}}  
  \def\MM{{\cal M}}

\def\dref#1{(\ref{#1})}

\def\be{\begin{equation}} \def\ee{\end{equation}}
\def\ba{\begin{array}} \def\ea{\end{array}} \def\bna{\begin{eqnarray}}
\def\ena{\end{eqnarray}}

 \def\NN{{\cal N}} \def\MM{{\cal M}}

\def\DD{{\cal D}}
\def\SS{{\cal S}}
 \def\XX{{\cal X}}   
\def\DD{{\cal D}}  
\def\TT{{\cal T}}

\def\YY{{\cal Y}}

 \def\bna{\begin{eqnarray}}
\def\ena{\end{eqnarray}} \def\dref#1{(\ref{#1})}
\begin{document}

\title{A Unified Relay Framework with\\ both D-F and C-F Relay Nodes}

\author{\authorblockN{Xiugang Wu and Liang-Liang Xie}\\
\authorblockA{
University of Waterloo, Waterloo, ON, Canada N2L 3G1 \\
Email: x23wu@uwaterloo.ca, llxie@uwaterloo.ca} }

\maketitle

\begin{abstract}
Decode-and-forward (D-F) and compress-and-forward (C-F) are two fundamentally different relay strategies proposed by (Cover and El Gamal, 1979). Individually, either of them has been successfully generalized to multi-relay channels. In this paper, to allow each relay node the freedom of choosing either of the two strategies, we propose a unified framework, where both the D-F and C-F strategies can be employed simultaneously in the network. It turns out that, to fully incorporate the advantages of both the best known D-F and C-F strategies into a unified framework, the major challenge arises as follows: For the D-F relay nodes to fully utilize the help of the C-F relay nodes, decoding at the D-F relay nodes should not be conducted until all the blocks have been finished; However, in the multi-level D-F strategy, the upstream nodes have to decode prior to the downstream nodes in order to help, which makes simultaneous decoding at all the D-F relay nodes after all the blocks have been finished inapplicable. To tackle this problem, nested blocks combined with backward decoding are used in our framework, so that the D-F relay nodes at different levels can perform backward decoding at different frequencies. As such, the upstream D-F relay nodes can decode before the downstream D-F relay nodes, and the use of backward decoding at each D-F relay node ensures the full exploitation of the help of both the other D-F relay nodes and the C-F relay nodes. The achievable rates under our unified relay framework are found to combine both the best known D-F and C-F achievable rates and include them as special cases.
\end{abstract}

\section{Introduction}
\begin{figure}[hbt]
\centering
\includegraphics[width=0.35\textwidth]{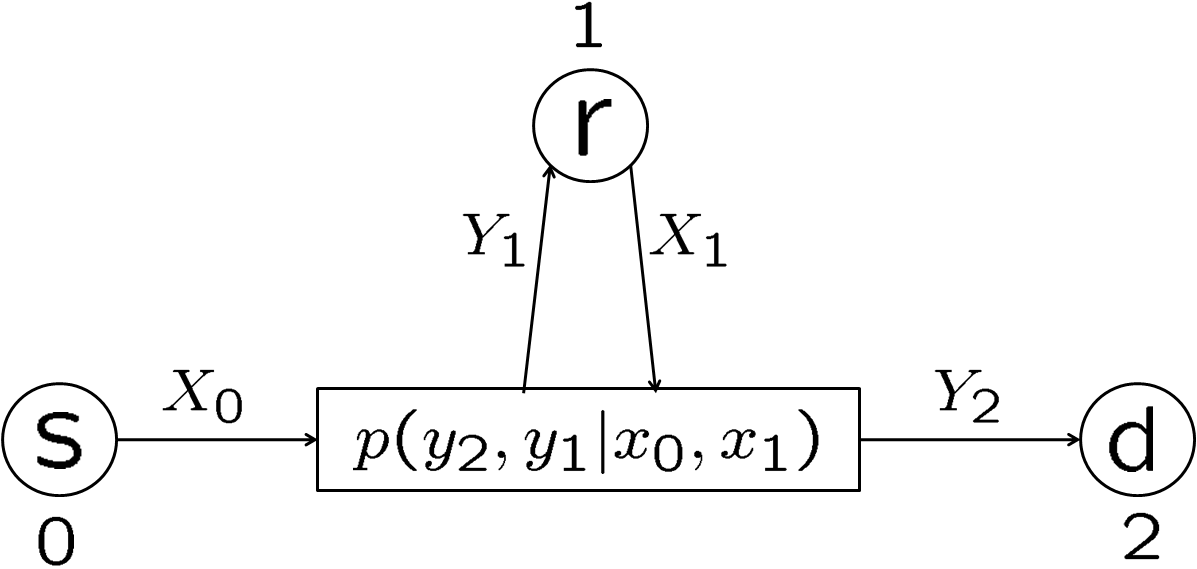}
\caption{The single-relay channel.}
 \label{single-relay}
\end{figure}

The relay channel, originally proposed in \cite{van71}, models a communication scenario where there is one or more relay nodes that can help the
information transmission between the source and the destination. The simplest one-relay channel is depicted
in Fig. \ref{single-relay}, where nodes 0, 1, and 2 are the source, the relay, and the destination, respectively. Two fundamentally different relay strategies have been
developed in \cite{covelg79} for such channels, which, depending on whether the relay decodes the information or not, are generally known
as {\it decode-and-forward} (D-F) and {\it compress-and-forward} (C-F)
respectively.

\subsection{D-F and C-F for single-relay channels}

In the D-F relay strategy, the relay first decodes the message sent by the source and then forwards it to the destination, and the destination decodes the message taking into account the inputs of both the source and the relay. With the D-F relay strategy, the following rate is achievable:
\begin{align}
R<\max_{p(x_0,x_1)}\min \{I(X_0;Y_1|X_1),I(X_0,X_1;Y_2)      \}\label{E:single-relayD-F}
\end{align}
where, the first condition $R<I(X_0;Y_1|X_1)$ makes node 1 able to decode the message based on the signal transmitted by node 0, and the second condition $R<I(X_0,X_1;Y_2)$ makes node 2 able to decode the message based on the signals transmitted by node 0 and node 1 together. Notably, the maximization in \dref{E:single-relayD-F} is over $p(x_0,x_1)$, rather than  $p(x_0)p(x_1)$, which suggests that \dref{E:single-relayD-F}
can only be achieved by node 0 and node 1 cooperating
with each other when transmitting signals.
%This is in fact counter-intuitive, since there is always a positive delay before
%node 1 can decode the information concerning the intention
%of node 0. But, by that time, node 0 would have moved on
%to transmit new information. Hence, node 1 can never catch up
%with node 0, which raises the issue of how they can cooperate
%together to transmit to node 2. Indeed, the D-F coding scheme developed in \cite{covelg79} to achieve \dref{E:single-relayD-F} is
%nontrivial, where, to tackle the above issue, an essential technique called block Markov coding was employed.
To accomplish such cooperation, an essential technique called block Markov coding was employed in the D-F coding scheme developed in \cite{covelg79}. Besides, the scheme  in \cite{covelg79} also used irregular encoding with codebooks of different sizes at the source and at the relay, random partitioning (binning), and successive decoding. Subsequently, some other D-F coding schemes also achieving \dref{E:single-relayD-F} were found in \cite{Carleial}-\cite{Willems}.

In contrast, the C-F relay strategy  is used when the
relay cannot decode the message sent by the source, but still can help
by compressing  its observation $Y_1$ into $\hat Y_1$, and forwarding this compressed version to the destination. The destination then either successively or jointly decodes the compression of the relay's observation and the original message of the source. In the original C-F scheme of  \cite{covelg79}, the decoder performs successive compression-message decoding, i.e., it first decodes the compression of the relay's observation, and then decodes the original message of the source, leading to the following achievable rate:
\begin{align}
R < &\max_{p(x_0)p(x_1)p(\hat y_1 |y_1,x_1)}I(X_0;\hat Y_1,Y_2|X_1) \label{cstr1}\\
\mbox{such that}~&~~I(Y_1;\hat Y_1|X_1,Y_2) \leq I(X_1;Y_2), \label{cstr2}
\end{align}
where  \dref{cstr2} ensures that the compression
$\hat Y_1$ can be first recovered at the
destination, and \dref{cstr1} ensures that the destination can decode the original message $X_0$ based on $\hat Y_1$ and $Y_2$ together.

The two-step compression-message successive decoding process in \cite{covelg79} requires $\hat Y_1$ to be decoded first, which facilitates the decoding of $X_0$, but is not a requirement of the original problem. Recognizing this, a joint compression-message decoding process was proposed in \cite{xie09}, where, instead of successively, the destination decodes $\hat Y_1$ and $X_0$ together. It turns out that the decoding of $X_0$ can be helped even if $\hat Y_1$ cannot be decoded first. In fact, with joint decoding, the constraint \dref{cstr2} is not necessary, and instead of \dref{cstr1}, the achievable rate is expressed as
\begin{equation}
\label{C-Fsinglerelay}
R<\max_{p(x_0)p(x_1)p(\hat y_1 |y_1,x_1)} I(X_0;\hat Y_1,Y|X_1)-\max\{0,I(Y_1;\hat Y_1|X_1,Y)-I(X_1;Y)\}. \footnote{Similar formulas as \dref{C-Fsinglerelay} have been derived with different arguments in \cite{ElGamal}-\cite{ElGamalKim}.}
\end{equation}
Therefore, compared to successive decoding, joint compression-message decoding provides more freedom in choosing the compression $\hat Y_1$. However, the question remains whether joint decoding achieves strictly higher rates for the original message than successive decoding. For the single relay case, it was proved in \cite{ElGamalKim} that the answer is negative, and any rate achievable by either of them can always be achieved by the other, i.e., the achievable rates in  \dref{cstr1}-\dref{cstr2} and \dref{C-Fsinglerelay} are essentially the same. In fact, as we will see later in the Introduction, when C-F is generalized to the case of multiple relays, there is no improvement on the achievable rate by joint decoding either.

Combining the D-F and C-F together, one can further consider the hybrid scheme, e.g., \cite[Thm 7]{covelg79}, where the relay partially decodes the message and compresses the rest of its received signals. However, such hybrid schemes generally involve superposition coding that induces auxiliary random variables, making the expression and evaluation of the achievable rates rather complicated especially in the case of multiple relays that we will consider in the sequel. Thus, in this paper, our discussion focuses on the ``pure'' D-F or C-F strategies only, i.e., the strategies where the relay either completely decodes the message, or does not decode at all but simply compresses and forwards its observation.

\subsection{D-F and C-F for multi-relay channels}
\begin{figure}[hbt]
\centering
\includegraphics[width=0.5\textwidth]{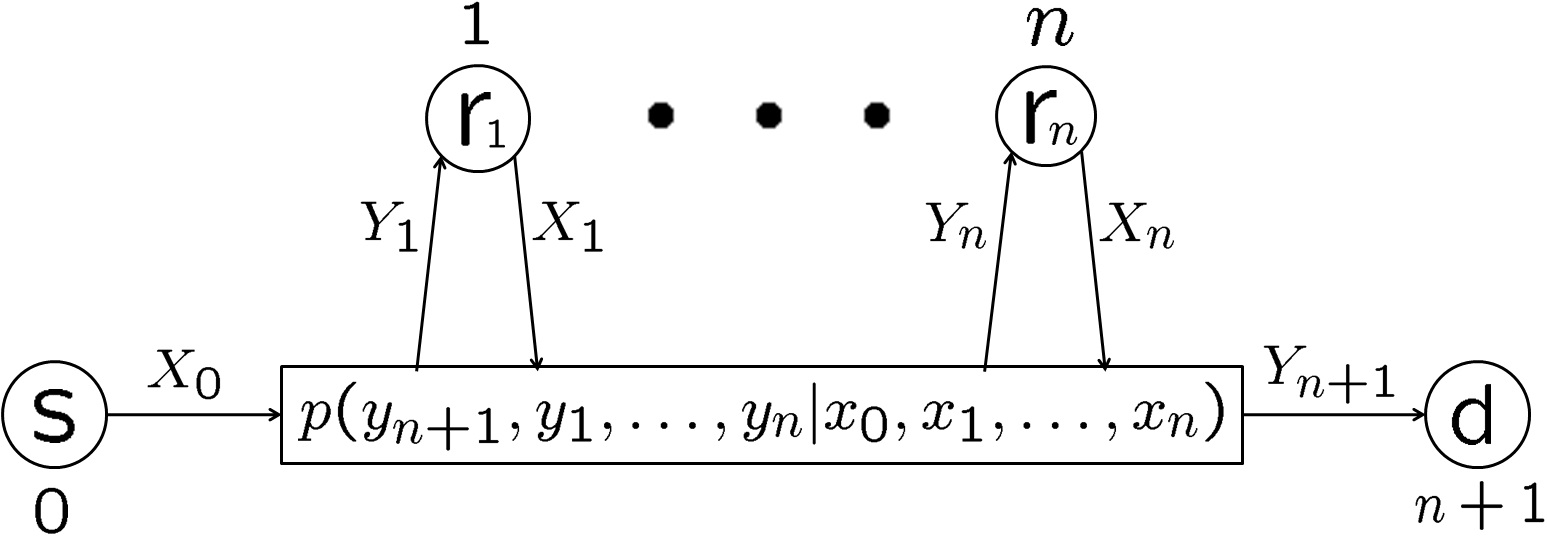}
\caption{The multiple-relay channel.}
\label{multi-relay}
\end{figure}

A natural extension of the single-relay channel in Fig. \ref{single-relay} is to the case of multiple relays depicted in Fig. \ref{multi-relay}, where nodes $0$ and $n+1$ are the source and the destination respectively, and nodes $1,2,\ldots, n$ are the $n$ relay nodes that constitute the relay nodes set, denoted by $\NN$. Both the D-F and C-F relay strategies have been separately generalized to such multi-relay channels in \cite{Aref}-\cite{WuXieTIT}, among which, \cite{XieKumar04}-\cite{kragasgup05} and \cite{KimElGamal}-\cite{WuXieTIT} provide the best achievable rates for D-F and C-F respectively.

Specifically, in generalizing D-F to the multi-relay channel, \cite{XieKumar04}-\cite{XieKumar05} modified the original irregular encoding/successive decoding scheme of \cite{covelg79} to a regular encoding/sliding window decoding scheme to realize the ``multi-level'' D-F relay strategy. For any fixed permutation $\pi$ on $\{0,1,\ldots,n+1\}$ with $\pi(1)=0$ and $\pi(n+2)=n+1$, i.e., any specific ordering of the relay nodes as $\pi(2),\pi(3),\ldots,\pi(n+1)$, their multi-level D-F scheme \cite{XieKumar04}-\cite{XieKumar05} achieves the following rate:
\begin{equation}
R<\max_{p(x_0,x_1,\ldots,x_n)}\min_{2 \leq k \leq n+2} I(X_{\pi(1:k-1)};Y_{\pi(k)}|X_{\pi(k:n+1)}), \label{multi-D-F}
\end{equation}
where $\pi(k_1:k_2):=\{\pi(k_1), \pi(k_1+1),\ldots,\pi(k_2) \}$. Later on, it was found in \cite{KramerAllerton03}-\cite{kragasgup05} that \dref{multi-D-F} can also be achieved with backward decoding.

The formula \dref{multi-D-F} has a similar interpretation as \dref{E:single-relayD-F}. For each
node $\pi(k),k=2,3,\ldots,n+2$, the corresponding rate constraint is
\begin{align}
R< I(X_{\pi(1:k-1)};Y_{\pi(k)}|X_{\pi(k:n+1)}),
\end{align}
which implies that for the decoding at node $\pi(k)$, the signals transmitted
by nodes $\pi(k+1:n+1)$ are known \emph{a priori}, and the
signals transmitted by nodes $\pi(1:k-1)$ are cooperating in providing
the information. A simple explanation of this feasibility is the following. In the multi-level D-F relay strategy, information is passed along the route $\pi(1) \rightarrow \pi(2) \rightarrow \cdots \rightarrow \pi(n+2)$, so that i) any information obtained by  the downstream nodes of $\pi(k)$, i.e., nodes $\pi(k+1:n+1)$, has already been obtained by node $\pi(k)$, and therefore their inputs are predictable by
node $\pi(k)$, and ii) by the time the information reaches node $\pi(k)$, all its upstream nodes $\pi(1:k-1)$ have already obtained the same information and can therefore
cooperate with the technique of block Markov coding. The formula \dref{multi-D-F} also demonstrates a remarkable feature of the multi-level D-F relay strategy in \cite{XieKumar04}-\cite{kragasgup05}, i.e., it completely eliminates the interference in the network: To any node, the signal transmitted by any other node is either a  ``real'' signal that can be used for decoding, or a \emph{priori} known signal that can be subtracted completely.

In the line of generalizing C-F to multi-relay channels, substantial advances have been recently made in \cite{KimElGamal}-\cite{WuXieTIT}. First,  in \cite{KimElGamal}, a new C-F scheme termed noisy network coding was proposed. Different from the original C-F scheme of \cite{covelg79}, where cumulative encoding/block-by-block forward decoding was used, this noisy network coding scheme employed repetitive encoding/all blocks united decoding. Besides, it also used compression-message joint decoding without uniquely decoding the relays' compressions, instead of compression-message successive decoding as in the original C-F scheme. It turns out \cite{KimElGamal} that the noisy network coding scheme achieves the same rate as the original C-F scheme for the single-relay channel, but improves the original C-F scheme in the case of multiple relays to achieve higher rates as follows:
\begin{align}
R<\max_{p(x_0)\prod_{i=1}^{n}p(x_i)p(\hat y_i|y_i,x_i)}\min_{\SS \subseteq \NN}I(X_0,X_\SS;\hat Y_{\NN \setminus \SS},Y_{n+1}|X_{\NN \setminus \SS})-I(Y_{\SS};\hat Y_{\SS}|X_0,X_\NN,Y_{n+1}, \hat Y_{\NN \setminus \SS}). \label{NNC}
\end{align}
%and for any $\SS\subseteq \NN$, $X_{\SS}$ denotes $\{X_i,i\in \SS\}$ and similarly for $Y_{\SS}$ and $\hat Y_{\SS}$.

However, soon in \cite{WuXieAllerton}-\cite{WuXieTIT}, it was discovered  that neither repetitive encoding/all blocks united decoding nor compression-message joint decoding used in \cite{KimElGamal}  is necessary to achieve the rate \dref{NNC}; in particular, a cumulative encoding/block-by-block backward decoding/compression-message successive decoding scheme was developed, and its corresponding achievable rate was shown to be the same as \dref{NNC}, with the following form:
\begin{align}
R< &\max_{p(x_0)\prod_{i=1}^{n}p(x_i)p(\hat y_i|x_i,y_i)} I(X_0;\hat Y_\NN, Y_{n+1}|X_\NN) \label{E:WuXie1} \\
\text{such that~} &I(X_\SS; \hat Y_{\NN \setminus \SS},Y_{n+1}|X_{\NN \setminus \SS}) -  I(Y_{\SS};\hat Y_{\SS}|X_\NN,Y_{n+1}, \hat Y_{\NN \setminus \SS})\geq 0 , \forall \SS \subseteq \NN, \label{E:WuXie2}
\end{align}
where \dref{E:WuXie1}-\dref{E:WuXie2} can be similarly interpreted as \dref{cstr1}-\dref{cstr2} for the single-relay case, i.e.,  \dref{E:WuXie2} ensures that the relays' compressions $\hat Y_{\NN}$ can be first recovered at the destination, and \dref{E:WuXie1} ensures that the destination can decode the original message of the source based on $Y_{n+1}$ and $\hat Y_{\NN}$ together. Note that the rate equivalence between \dref{NNC} and \dref{E:WuXie1}-\dref{E:WuXie2} also demonstrates that in the case of multiple relays, there is no improvement on the achievable rate by joint compression-message decoding either, which is consistent with the conclusion made in the single-relay case. More interestingly, in proving such a rate equivalence,  \cite{WuXieAllerton}-\cite{WuXieTIT} found that the the R.H.S. (right-hand-side) of \dref{NNC} is maximized \emph{only when} the compressions $\hat Y_{\NN}$  are chosen to satisfy \dref{E:WuXie2}, i.e., to maximize the achievable rate for the original message, the compressions should always be chosen to support successive decoding, and any compressions not supporting successive decoding will actually lead to strictly lower achievable rates for the original message.

Since block-by-block backward decoding and compression-message successive decoding are relatively easier to implement than all blocks united decoding and compression-message joint decoding respectively, the  cumulative encoding/block-by-block backward decoding/compression-message successive decoding scheme of \cite{WuXieAllerton}-\cite{WuXieTIT} becomes the simplest choice in achieving the highest C-F rate in the case of multiple relays. Moreover, the fact that this scheme achieves the same rate as noisy network coding also reveals the essential reason for the improvement of the achievable rate -- not repetitive encoding/all blocks united decoding, not joint compression-message decoding, but delayed decoding until all the blocks have been finished. This delayed decoding is generally necessary because the multiple-relay case differs from the single-relay case in that it may take multiple blocks for the relays to help each
other before their compressions can finally reach the destination. Hence, the block-by-block forward
decoding scheme, which is sufficient for the single-relay case, may not work satisfactorily for multiple
relays in general \cite{WuXieTIT}.

It is worth noting that although the optimal C-F rate is achieved only when the compressions are chosen to support successive decoding in single-destination networks, in a network with multiple destinations (\cite{KimElGamal}, \cite{WuXieTIT}), the compressions may not be chosen to support successive decoding at all the destinations, and joint decoding might have to be used. For this, a more general scheme of cumulative encoding/block-by-block backward decoding/compression-message joint decoding was developed in \cite{WuXieTIT}. For any given distribution $p(x_0)\prod_{i=1}^{n}p(x_i)p(\hat y_i|x_i,y_i)$, this scheme achieves the following rate:
\begin{equation}
\label{cbjrate}
R<\min_{\SS \subseteq \DD}I(X_0,X_{\SS};\hat Y_{\DD \setminus \SS},Y_{n+1}|X_{\DD \setminus \SS})-I(Y_{\SS};\hat Y_{\SS}|X_0,X_{\DD},Y_{n+1}, \hat Y_{\DD \setminus \SS}),
\end{equation}
where $\DD$ is the unique largest subset of $\NN$ satisfying
\begin{equation}
\label{eq1}
I(X_\SS;\hat Y_{\DD \setminus \SS},Y_{n+1}|X_0,X_{\DD \setminus \SS})-I(Y_{\SS};\hat Y_{\SS}|X_0,X_{\DD},Y_{n+1}, \hat Y_{\DD \setminus \SS})> 0,  \forall \SS \subseteq \DD, \SS \neq \emptyset,
\end{equation}
and $\hat Y_{\DD}$ can be decoded jointly with $X_0$. Here, $\DD$ can be interpreted as the ``jointly decodable'' relay nodes set such that the compressions of the relays in this set are decodable jointly with the original message $X_0$. In contrast, the compression of any relay node  in $\NN \setminus \DD'$ is not decodable even jointly with $X_0$, where $\DD'$ is the unique largest subset of $\NN$ satisfying
\begin{equation}
\label{eq2}
I(X_\SS;\hat Y_{\DD'  \setminus \SS},Y_{n+1}|X_0,X_{\DD'  \setminus \SS})-I(Y_{\SS};\hat Y_{\SS}|X_0,X_{\DD'},Y_{n+1}, \hat Y_{\DD' \setminus \SS})\geq 0, \forall \SS \subseteq \DD'.
\end{equation}

On the other hand, for any given distribution $p(x_0)\prod_{i=1}^{n}p(x_i)p(\hat y_i|x_i,y_i)$, the achievable rate \dref{NNC} can be more generally expressed as
\begin{equation}
\label{NNCimproved}
R<\min_{\SS \subseteq \TT}I(X_0,X_\SS;\hat Y_{\TT \setminus \SS},Y_{n+1}|X_{\TT \setminus \SS})-I(Y_{\SS};\hat Y_{\SS}|X_0,X_\TT,Y_{n+1}, \hat Y_{\TT \setminus \SS})
\end{equation}
if we only consider a subset of relays $\TT\subseteq \NN$ for the decoding, while treating the other inputs as purely noise. Interestingly, it was found in \cite{WuXieTIT} that
among all the choices of $\TT \subseteq \NN$, the R.H.S. of \dref{NNCimproved} is maximized when $\TT=\DD$ or $\TT=\DD'$, but is strictly less than the maximum when $\TT \nsubseteq \DD'$. Therefore, only those relays whose compressions are jointly decodable are helpful to the decoding of the original message, and including the jointly un-decodable compressions in the formula \dref{NNCimproved}, i.e., choosing $\TT \nsubseteq \DD' $, will even strictly lower the achievable rate.

By comparing \dref{cbjrate} and \dref{NNCimproved} with $\TT=\DD$, it can be concluded that for any compressions chosen at the relays, the cumulative encoding/block-by-block backward decoding/compression-message joint decoding scheme of \cite{WuXieTIT} achieves the same rate as the noisy network coding scheme \cite{KimElGamal}.\footnote{Part of the results in \cite{WuXieAllerton}-\cite{WuXieTIT} have also been recognized in \cite{KramerHou11}-\cite{KramerHou11ITW}.}
%Up to now, at least three coding schemes have been developed that are capable of achieving (2). (The most recent survey on relay channels can be found in [5].) The original coding
%scheme designed in [3] uses several complex techniques: block Markov superposition encoding, random partitioning (binning), and successive decoding. This scheme even uses codebooks of different sizes. Later on, a much simpler coding scheme was developed by Carleial [6] in the study of multiple-access channel with generalized feedback (MAC-GF), which includes the one-relay channel as a special case. This new scheme still uses block Markov superposition encoding, but avoids random partitioning, and all codebooks are of the same size. The key new idea lies in the decoding: Unlike the sequential manner in [3], it is a simultaneous typicality check of two consecutive blocks. But the paper [6] itself did not point out that this was a new scheme for achieving (2). The third scheme achieving (2) is the backward decoding introduced in [7]. When MAC-GF is concerned, the backward decoding is more powerful in achieving higher rates than Carleial¡¯s scheme as was shown in [8]. But for relay channels, they achieve the same rates. Moreover, since the backward decoding starts the decoding process only after all the blocks have been received, it incurs a substantial decoding delay.
\subsection{A unified relay framework with both D-F and C-F relay nodes}
In the above discussions, all the relay nodes in the network perform only one type of relay strategy, either D-F or C-F. However, to obtain higher achievable rate, it might be better to let each relay node choose from D-F and C-F its relay strategy depending on the channel condition, e.g., let the relay node close to the source perform D-F while let the relay node close to the destination perform C-F. This invokes a unified relay framework that includes both the D-F and C-F relay nodes in the network. In developing such a framework, one naturally wants to combine the advantages of
both the best known D-F and C-F schemes, i.e., the multi-level D-F schemes in \cite{XieKumar04}-\cite{kragasgup05} and the recent advances on C-F schemes in \cite{KimElGamal}-\cite{WuXieTIT}.

%There have been some investigations on developing the above mentioned unified relay framework,  but they are all done before the recent advance in the C-F relay strategy [], and thus have not taken into account the recent advance in the C-F relay strategy.
%
% However, it failed to utilize the
An attempt towards this unified relay framework has been recently made in \cite{ISIT12}. In the scheme of \cite{ISIT12}, part of the relay nodes use D-F and the rest use C-F, and the D-F relay nodes exploit the help of the C-F relay nodes via offset coding. However, the scheme in \cite{ISIT12} failed to take full advantage of the best known D-F and C-F strategies. Firstly, \cite{ISIT12} didn't use the multi-level D-F schemes as in \cite{XieKumar04}-\cite{kragasgup05}. Instead, all the D-F relay nodes in the scheme of \cite{ISIT12} are at the same level, and thus the decoding at each D-F relay node couldn't exploit the help of other D-F relay nodes. Secondly,  in \cite{ISIT12}, although the destination performed backward decoding to fully exploit the help of the C-F relay nodes, the decoding at each D-F relay node was based on two consecutive blocks only and thus didn't fully utilize the help of the C-F relay nodes as in \cite{KimElGamal}-\cite{WuXieTIT}. (Note as mentioned in Part B, in the case of multiple C-F relay nodes, delayed decoding after all the blocks have been finished is in general necessary.)

Indeed, it turns out that, to fully incorporate the advantages of both the best known D-F and C-F relay strategies into a unified framework is nontrivial due to the following major challenge: For the D-F relay nodes to fully utilize the help of the C-F relay nodes as in \cite{KimElGamal}-\cite{WuXieTIT}, decoding at the D-F relay nodes
should not be conducted until all the blocks have been finished; However, to perform the multi-level D-F strategy as in \cite{XieKumar04}-\cite{kragasgup05}, the
upstream nodes have to decode prior to the downstream nodes in order to help, which makes simultaneous decoding
at all the D-F relay nodes after all the blocks have been finished inapplicable.

To tackle this problem, nested blocks (\cite{KramerAllerton03}-\cite{kragasgup05}, \cite{XieKumar07}) combined with backward decoding
are used in our framework, so that the D-F relay nodes at different levels can perform backward decoding at different frequencies: the closer to the source in the information passing route, the higher decoding frequency.  As such, the upstream D-F relay nodes can decode before the downstream D-F relay nodes and the destination, and the use of backward decoding at each D-F relay node ensures the full exploitation of the help of both the other D-F relay nodes and the C-F relay nodes.

\begin{figure}[hbt]
\centering
\includegraphics[width=0.5\textwidth]{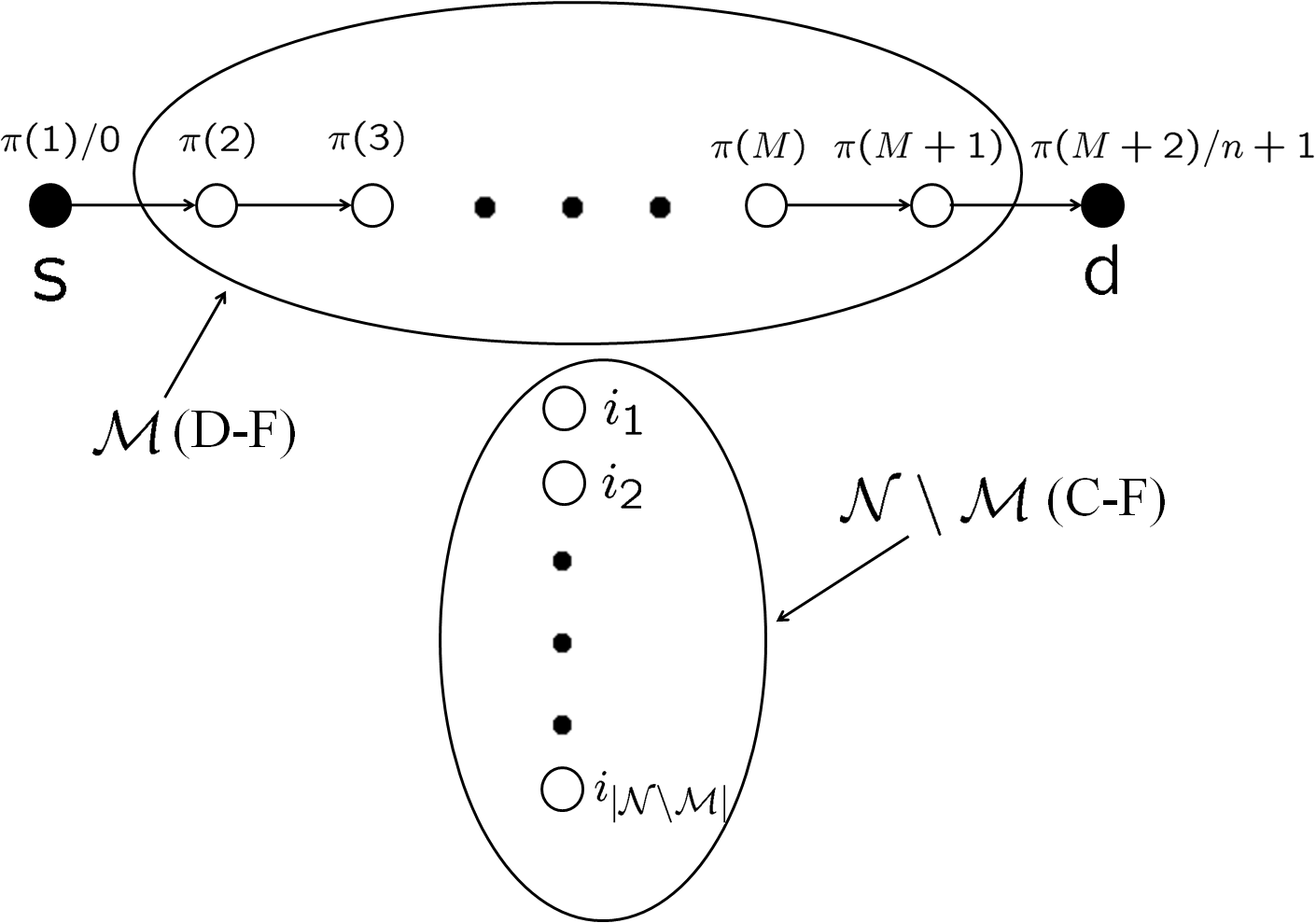}
\caption{A unified relay framework with both the D-F and C-F relay nodes.}
 \label{DF-CF}
\end{figure}

Specifically, we partition the relay nodes set $\NN$ into two sets, $\MM$ with $|\MM|=M$ and $\NN \setminus \MM$, as depicted in Fig. \ref{DF-CF}, and fix some permutation $\pi$ on $\{0\}\bigcup \MM \bigcup \{n+1\}$ with $\pi(1)=0$ and $\pi(M+2)=n+1$. Let the relay nodes in $\MM$ perform the multi-level D-F cooperatively along the route  $\pi(1) \rightarrow \pi(2) \rightarrow \cdots \rightarrow \pi(M+2)$, while let each node $i\in \NN \setminus \MM$ performs C-F as in \cite{KimElGamal}-\cite{WuXieTIT} independently. Then, a total of $B^{M+1}$ blocks will be used and the length of a ``virtual'' block for node $\pi(k),k=2,3,\ldots,M+2$, will be $B^{k-2}$. The backward decoding at the destination, i.e., node $\pi(M+2)$, will happen at the end of all $B^{M+1}$ blocks, while the backward decoding at the D-F relay node $\pi(k), k=2,3,\ldots, M+1$, will happen whenever it has received $B$ new ``virtual'' blocks, i.e., at the end of each block $b =vB^{k-1},  v\in [1:B^{M+1}/B^{k-1}]$. Also, both the D-F relay nodes and the destination will perform compression-message joint decoding, which is in general necessary since the compressions of the C-F relay nodes may not be chosen to support successive decoding at all the D-F relay nodes and the destination.

Under the above described framework, for any given distribution $p(x_0)p(x_{\MM}|x_0)\prod_{i\in \NN \setminus \MM}p(x_i)p(\hat y_i|y_i,x_i)$, the following rate is achievable:
\begin{equation}
R<\min_{2 \leq k \leq M+2} \min_{\SS \subseteq \DD_{k}}I(X_{\pi(1:k-1)},X_\SS;\hat Y_{\DD_{k} \setminus \SS},Y_{\pi(k)}|X_{\DD_{k} \setminus \SS},X_{\pi(k:M+1)})-I(Y_{\SS};\hat Y_{\SS}|X_{\pi(1:M+1)},X_{\DD_{k}},Y_{\pi(k)}, \hat Y_{\DD_{k} \setminus \SS}), \label{E:Intro1}
\end{equation}
where $\DD_{k}$ is the unique largest subset of $\NN \setminus \MM$ satisfying
\begin{equation}
I(X_\SS;\hat Y_{\DD_{k} \setminus \SS},Y_{\pi(k)}|X_{\pi(1:M+1)},X_{\DD_{k} \setminus \SS})-I(Y_{\SS};\hat Y_{\SS}|X_{\pi(1:M+1)},X_{\DD_{k}},Y_{\pi(k)}, \hat Y_{\DD_{k} \setminus \SS})> 0, \label{E:Intro2}
\end{equation}
for any nonempty $\SS \subseteq \DD_{k}$.

\dref{E:Intro1} has the flavors of both \dref{multi-D-F} and  \dref{cbjrate}. Specifically, for each node $\pi(k),  k=2,3,\ldots, M+2$, the corresponding rate constraint is
\begin{equation}
R< \min_{\SS \subseteq \DD_{k}}I(X_{\pi(1:k-1)},X_\SS;\hat Y_{\DD_{k} \setminus \SS},Y_{\pi(k)}|X_{\DD_{k} \setminus \SS},X_{\pi(k:M+1)})-I(Y_{\SS};\hat Y_{\SS}|X_{\pi(1:M+1)},X_{\DD_{k}},Y_{\pi(k)}, \hat Y_{\DD_{k} \setminus \SS}), \label{E:1analysis}
\end{equation}
which is in a form similar to \dref{cbjrate} but with the appearance of $X_{\pi(1:k-1)}$, $X_{\pi(k:M+1)}$ and $X_{\pi(1:M+1)}$.  \dref{E:1analysis} has the similar form as \dref{cbjrate} since node $\pi(k)$ uses the help of the C-F relay nodes as in \cite{WuXieAllerton}-\cite{WuXieTIT}. $X_{\pi(1:k-1)}$, $X_{\pi(k:M+1)}$ and $X_{\pi(1:M+1)}$ appear in \dref{E:1analysis} because node $\pi(k)$ also utilizes the help of other D-F relay nodes as in  \cite{XieKumar04}-\cite{kragasgup05} so that the signals of its upstream nodes, i.e., $X_{\pi(1:k-1)}$, are cooperatively providing the information while the signals of its downstream nodes and itself $X_{\pi(k:M+1)}$ are known at $\pi(k)$.  Also, the set $\DD_{k}$ defined in \dref{E:Intro2} has a similar interpretation as the set $\DD$ defined in \dref{eq1}, i.e., the ``jointly decodable'' C-F relay nodes set at node $\pi(k)$ such that the compressions of the relays in this set are decodable jointly with $X_{\pi(1:k-1)}$ given that $X_{\pi(k:M+1)}$ are known at node $\pi(k)$.

It can be easily seen that \dref{E:Intro1} includes the achievable rates in \dref{multi-D-F} and  \dref{cbjrate} as special cases: When $\MM=\NN$, i.e., all the relays perform D-F, $\DD_{k}=\emptyset$ and \dref{E:Intro1} reduces to  \dref{multi-D-F};
When $\MM=\emptyset$, i.e., all the relays perform C-F, \dref{E:Intro1} reduces to \dref{cbjrate}.

Finally, it should be noted that, the achievable rate \dref{E:Intro1} is proved by using the block-by-block backward decoding scheme in \cite{WuXieTIT}.
We can also modify the all blocks united decoding scheme in \cite{KimElGamal} to a $B$-blocks-by-$B$-blocks backward decoding scheme, to fit it into our unified relay framework and prove the following achievable rate:
\begin{equation}
\label{NNCbased}
R<\min_{2 \leq k \leq M+2}\max_{\TT_k \subseteq \NN \setminus \MM} \min_{\SS \subseteq \TT_k} I(X_{\pi(1:k-1)},X_\SS;\hat Y_{\TT_k \setminus \SS},Y_{\pi(k)}|X_{\TT_k \setminus \SS},X_{\pi(k:M+1)})-I(Y_{\SS};\hat Y_{\SS}|X_{\pi(1:M+1)},X_{\TT_k},Y_{\pi(k)}, \hat Y_{\TT_k \setminus \SS}).
\end{equation}
Similarly to the equivalence between \dref{cbjrate} and \dref{NNCimproved}, here \dref{E:Intro1} and \dref{NNCbased} are also equivalent. One can also easily check that \dref{NNCbased} includes the achievable rates in \dref{multi-D-F} and  \dref{NNCimproved} as special cases by letting $\MM=\NN$ and $\MM=\emptyset$ respectively. Notably, again, in terms of complexity, block-by-block backward decoding is relatively easier to implement since $B$-blocks-by-$B$-blocks backward decoding involves $B$ blocks united decoding.

The remainder of the paper is organized as the following. In Section \ref{S:mainresults}, we formally state our problem setup and summarize the main results. Then, in Section \ref{S:byb} and Section \ref{S:Bblocks}, our unified relay framework with block-by-block backward decoding and with $B$-blocks-by-$B$-blocks backward decoding will be presented in detail respectively. Finally, some concluding remarks are included in Section \ref{conclusion}.

\section{Main Results}\label{S:mainresults}
Consider a multiple-relay channel consisting of $n+2$ nodes, as depicted in Fig. \ref{multi-relay}, where nodes $0$ and $n+1$ are the source and the destination respectively, and nodes $1,2,\ldots, n$ are the $n$ relay nodes. Formally, this channel can be denoted by
$$
(\XX_0 \times \XX_1\times\cdots\times \XX_n,\,p(y_{n+1},y_1,\ldots,y_n|x_0,x_1,\ldots,x_n),\,\YY_{n+1} \times \YY_1\times \cdots \times \YY_n)
$$
where, $\XX_0,\XX_1,\ldots,\XX_n$
are the transmitter alphabets of the source and the relays respectively,
$\YY_{n+1},\YY_1,\ldots,\YY_n$ are the receiver alphabets of the destination and
the relays respectively, and a collection of probability distributions
$p(\cdot,\cdot,\ldots,\cdot|x_0,x_1,\ldots,x_n)$ on $\YY_{n+1} \times \YY_1\times \cdots \times \YY_n$, one for each
$(x_0,x_1,\ldots,x_n)\in \XX_0 \times \XX_1\times\cdots\times \XX_n$. The interpretation is that $x_0$
is the input to the channel from the source, $y_{n+1}$ is
the output of the channel to the destination, and $y_i$ is
the output received by the $i$-th relay. The $i$-th relay sends an input $x_i$ based on what it has received:
\begin{equation}
\label{processing2}
x_i(t)=r_{i,t}(y_i(t-1),y_i(t-2),\ldots), ~~ \mbox{ for every time } t,
\end{equation}
where $r_{i,t}(\cdot)$ can be any causal function.

Before presenting the main results, we introduce some simplified notations. Denote the set $\NN=\{1,2,\ldots,n\}$. For any subset $\SS\subseteq \{0, 1, \ldots, n+1\}$, let $X_{\SS}=\{X_i,i\in \SS\}$, and use similar notations for other variables. For any $\MM \subseteq \NN$ with $|\MM|=M$, let $\pi(\{0,\MM,n+1\})$ be a permutation on $\{0\}\bigcup \MM \bigcup \{n+1\}$ with $\pi(1)=0$ and $\pi(M+2)=n+1$, and let $\pi(k_1:k_2)=\{\pi(k_1), \pi(k_1+1),\ldots,\pi(k_2) \}$.

Under our unified relay framework as described in the Introduction, the following Theorems \ref{main1} and  \ref{main2} present the achievable rates by block-by-block backward decoding and $B$-blocks-by-$B$-blocks backward decoding respectively. The coding schemes used to prove these theorems constitute the key contributions of our paper, and will be presented in detail in Sections \ref{S:byb} and \ref{S:Bblocks} respectively.
\begin{theorem}\label{main1}
For the multiple-relay channel, a rate $R$ is achievable if for some $\MM \subseteq \NN$ with $|\MM|=M$, there exists a permutation $\pi(\{0,\MM,n+1\})$ and some
$$
p(q)p(x_0|q)p(x_{\MM}|x_0,q)\prod_{i\in \NN \setminus \MM}p(x_i|q)p(\hat y_i|y_i,x_i,q),
$$
such that for any $k=2,3,\ldots, M+2$,
\begin{equation}
R<\min_{\SS \subseteq \DD_{k}}I(X_{\pi(1:k-1)},X_\SS;\hat Y_{\DD_{k} \setminus \SS},Y_{\pi(k)}|X_{\DD_{k} \setminus \SS},X_{\pi(k:M+1)},Q)-I(Y_{\SS};\hat Y_{\SS}|X_{\pi(1:M+1)},X_{\DD_{k}},Y_{\pi(k)}, \hat Y_{\DD_{k} \setminus \SS},Q), \label{E:constraint1}
\end{equation}
where $\DD_{k}$ is the unique largest subset of $\NN \setminus \MM$ satisfying
\begin{equation}
I(X_\SS;\hat Y_{\DD_{k} \setminus \SS},Y_{\pi(k)}|X_{\pi(1:M+1)},X_{\DD_{k} \setminus \SS},Q)-I(Y_{\SS};\hat Y_{\SS}|X_{\pi(1:M+1)},X_{\DD_{k}},Y_{\pi(k)}, \hat Y_{\DD_{k} \setminus \SS},Q)> 0, \label{E:constraint2}
\end{equation}
for any nonempty $\SS \subseteq \DD_{k}$.
\end{theorem}

\begin{theorem}\label{main2}
For the multiple-relay channel, a rate $R$ is achievable if for some $\MM \subseteq \NN$ with $|\MM|=M$, there exists a permutation $\pi(\{0,\MM,n+1\})$ and some
$$
p(q)p(x_0|q)p(x_{\MM}|x_0,q)\prod_{i\in \NN \setminus \MM}p(x_i|q)p(\hat y_i|y_i,x_i,q),
$$
such that for any $k=2,3,\ldots, M+2$,
\begin{equation}
\label{thmNNCbased}
R<\max_{\TT_k \subseteq \NN \setminus \MM} \min_{\SS \subseteq \TT_k} I(X_{\pi(1:k-1)},X_\SS;\hat Y_{\TT_k \setminus \SS},Y_{\pi(k)}|X_{\TT_k \setminus \SS},X_{\pi(k:M+1)},Q)-I(Y_{\SS};\hat Y_{\SS}|X_{\pi(1:M+1)},X_{\TT_k},Y_{\pi(k)}, \hat Y_{\TT_k \setminus \SS},Q).
\end{equation}
\end{theorem}

The following theorem establishes the equivalence between the achievable rates in Theorems  \ref{main1} and \ref{main2}. The proof of this theorem can be immediately obtained by analogy to the proof of \cite[Thm 2.8]{WuXieTIT} and will be omitted in this paper.
\begin{theorem}
\label{T:equi}
For any $\MM \subseteq \NN$ with $|\MM|=M$, any permutation $\pi(\{0,\MM,n+1\})$, any distribution $$p(q)p(x_0|q)p(x_{\MM}|x_0,q)\prod_{i\in \NN \setminus \MM}p(x_i|q)p(\hat y_i|y_i,x_i,q),$$
and any $k=2,3,\ldots,M+2$, the maximum in the R.H.S. of \dref{thmNNCbased}  is attained when $\TT_k=\DD_{k}$, where $\DD_{k}$  is as defined in \dref{E:constraint2}.
\end{theorem}

\begin{remark}
Finally, we point out that Theorems \ref{main1} and \ref{main2} can also be applied to multiple-destination problems, by choosing the D-F relay nodes set $\MM$ to include the other destinations.
\end{remark}

\section{Unified Relay Framework With Block-By-Block Backward Decoding}\label{S:byb}
To prove Theorem \ref{main1}, we incorporate the multi-level D-F scheme in \cite{XieKumar04}-\cite{kragasgup05} and the cumulative encoding/block-by-block backward decoding/comression-message joint decoding C-F scheme in \cite{WuXieTIT} into the unified relay framework described in the Introduction.

Specifically, we divide the relay set $\NN$ into two sets, $\MM$ with $|\MM|=M$ and $\NN \setminus \MM$, as shown in Fig. \ref{DF-CF}, and fix some permutation $\pi(\{0,\MM,n+1\})$ with $\pi(1)=0$ and $\pi(M+2)=n+1$. The source performs cumulative encoding, in the sense that a new message is encoded at the source in each new block; the nodes in $\MM$ perform multi-level D-F cooperatively, along the route $\pi(1) \rightarrow \pi(2) \rightarrow \cdots \rightarrow \pi(M+2)$, in a similar manner with \cite{XieKumar04}-\cite{kragasgup05}; each node $i\in \NN \setminus \MM$ performs C-F independently in the same way as \cite{KimElGamal}-\cite{WuXieTIT}; both the D-F relay nodes and the destination node, i.e., nodes $\pi(2:M+2)$, perform  compression-message joint decoding in a block-by-block backward manner. (Note here, the nodes $\pi(2:M+2)$ will be treated as multiple destinations with respect to the C-F relay nodes, and thus compression-message joint decoding is generally necessary at the these nodes, as mentioned in the Introduction.)
A total of $B^{M+1}$ blocks will be used and the length of a ``virtual'' block for node $\pi(k),k=2,3\ldots,M+2$, will be $B^{k-2}$. The backward decoding at the destination, i.e., node $\pi(M+2)$, will happen at the end of all $B^{M+1}$ blocks, while the backward decoding at the D-F relay node $\pi(k), k=2,3,\ldots, M+1$, will happen at the end of every $B^{k-1}$ blocks, i.e., at the end of block $b =vB^{k-1},  v\in [1:B^{M+1}/B^{k-1}]$.

To make the presentation of the detailed coding scheme easier to follow, we first consider the case of single D-F relay node, i.e., when $M=1$, and then present the extension to the general case of multiple D-F relay nodes, i.e., when $M\geq 2$.

\subsection{Single D-F relay node ($M=1$)}
Assume that, among the relay nodes set $\NN$, only node 1 is the D-F relay node, and all other relay nodes are the C-F relay nodes. Denote $\tilde{\NN}=\NN \setminus \{1\}$. Specializing Theorem  \ref{main1} to this case, we have that a rate $R$ is achievable, if there exists  some $$p(q)p(x_0|q)p(x_1|x_0,q)\prod_{i\in \tilde{\NN}}p(x_i|q)p(\hat y_i|y_i,x_i,q),$$
such that
\begin{equation}
R<\min \left\{
\begin{split}
&\min_{\SS \subseteq \DD_{1}}  I(X_0,X_\SS;\hat Y_{\DD_{1} \setminus \SS},Y_{1}|X_1, X_{\DD_{1} \setminus \SS},Q)-I(Y_{\SS};\hat Y_{\SS}|X_{0},X_1, X_{\DD_1},Y_{1}, \hat Y_{\DD_{1} \setminus \SS},Q) \\
&\min_{\SS \subseteq \DD_{n+2}}  I(X_0,X_1,X_\SS;\hat Y_{\DD_{n+2} \setminus \SS},Y_{n+2}|X_{\DD_{n+2} \setminus \SS},Q)-I(Y_{\SS};\hat Y_{\SS}|X_{0},X_1, X_{\DD_{n+2}},Y_{n+2}, \hat Y_{\DD_{n+2} \setminus \SS},Q)
\end{split}
\right\}\label{E:simplified}
\end{equation}
where $\DD_{1}$ is the unique largest subset of $\tilde{\NN}$ satisfying
\begin{equation}
I(X_\SS;\hat Y_{\DD_{1} \setminus \SS},Y_{1}|X_{0},X_1, X_{\DD_{1} \setminus \SS},Q)-I(Y_{\SS};\hat Y_{\SS}|X_{0},X_1,X_{\DD_{1}},Y_{1}, \hat Y_{\DD_{1} \setminus \SS},Q)> 0,\label{E:sim1}
\end{equation}
for any nonempty $\SS \subseteq \DD_{1}$, and $\DD_{n+2}$ is the unique largest subset of $\tilde{\NN}$ satisfying
\begin{equation}
I(X_\SS;\hat Y_{\DD_{n+2} \setminus \SS},Y_{n+2}|X_{0},X_1, X_{\DD_{n+2} \setminus \SS},Q)-I(Y_{\SS};\hat Y_{\SS}|X_{0},X_1,X_{\DD_{n+2}},Y_{n+2}, \hat Y_{\DD_{n+2} \setminus \SS},Q)> 0,\label{E:sim2}
\end{equation}
for any nonempty $\SS \subseteq \DD_{n+2}$.

The uniqueness of  $\DD_{1}$ and $\DD_{n+2}$ can be immediately obtained by analogy to the proof of \cite[Thm 2.7]{WuXieTIT}. Below, we focus on proving the achievablity of the rate in \dref{E:simplified}. For simplicity of notation, we only prove the achievability for the case $Q=\emptyset$. Achievability for an arbitrary time-sharing random variable $Q$ can be obtained by using the standard technique of time sharing \cite{CoverThomas}, \cite{ElGamalKim}. The same consideration on $Q$ applies throughout all the proofs of this paper.

In the case of single D-F relay node, a total of $B^{2}$ blocks will be used. The backward decoding at the destination node $n+2$ will happen at the end of all $B^{2}$ blocks, while the backward decoding at the D-F relay node $1$ will happen at the end of every $B$ blocks, i.e., at the end of block $b =vB,  v\in [1:B]$. Note here, in order to fully utilize the help of the C-F nodes as in \cite{KimElGamal}-\cite{WuXieTIT}, even the only D-F relay node $1$, has to perform backward decoding, which is different from the situation arising in \cite{KramerAllerton03}-\cite{kragasgup05} and \cite{XieKumar07}, where there is no issue of exploiting the help of the C-F nodes and node $1$ can decode at the end of every block. The detailed codebook generation and encoding/decoding process are as follows, which can be understood with the help of Table \ref{tab}.

\emph{Codebook Generation:} Fix $p(x_0)p(x_1|x_0)\prod_{i\in\tilde{ \NN} }p(x_i)p(\hat y_i|y_i,x_i)$. We randomly and independently generate a codebook for each block.

i) First consider the codebook generation for the source node 0 and the D-F relay node 1. A joint codebook for these two nodes will be generated in a backward manner similar to \cite{XieKumar05} for each block. Specifically, for each block $b\in [1: B^{2}]$,  randomly generate $2^{TR}$ independent sequences $\mathbf{x}_{1,b}(m_{b-B})$ for node 1, and randomly generate $2^{TR}$ conditionally independent sequences $\mathbf{x}_{0,b}(m_{b}|m_{b-B})$ for node 0, where $m_{b},m_{b-B}\in [1:2^{TR}]$. As in \cite{XieKumar05}, the codebook is generated in the backward manner because the source node 0 knows what the D-F relay node 1 is going to transmit, and therefore can adjust its own transmission accordingly, but not the converse. The difference from \cite{XieKumar05} is that here the delay between the messages transmitted by node 1 and node 0 is $B$ blocks, instead of 1 block in \cite{XieKumar05}, since in our framework node 1 has to wait for every $B$ blocks to perform backward decoding for exploiting the help of the C-F relay nodes.

ii) Then we generate the codebooks for the C-F relay nodes in the same way as in \cite{KimElGamal}-\cite{WuXieTIT}. For each block $b\in [1: B^{2}]$ and each relay node $i\in \tilde{\NN}$, randomly and independently generate $2^{T\hat R_i}$ sequences $\mathbf{x}_{i,b}(l_{i,b-1})$, $l_{i,b-1}\in [1:2^{T\hat R_i}]$, where $\hat R_i=I(Y_i;\hat Y_i|X_i)+\epsilon$;
for each relay node $i\in \tilde{\NN}$ and each $\mathbf{x}_{i,b}(l_{i,b-1})$, $l_{i,b-1}\in [1:2^{T\hat R_i}]$, randomly and conditionally independently generate $2^{T\hat R_i}$ sequences $ \hat{\mathbf{y}}_{i,b}(l_{i,b}|l_{i,b-1})$, $l_{i,b}\in [1:2^{T\hat R_i}]$.

The combination of i) and ii) defines the codebook for any block $b\in [1: B^{2}]$,
\begin{align}
\mathcal{C}_b=\Big\{&\mathbf{x}_{1,b}(m_{b-B}), \mathbf{x}_{0,b}(m_{b}|m_{b-B}): m_{b},m_{b-B}\in [1:2^{TR}];  \nonumber \\
 &\mathbf{x}_{i,b}(l_{i,b-1}) , \hat{\mathbf{y}}_{i,b}(l_{i,b}|l_{i,b-1}):  l_{i,b},l_{i,b-1} \in  [1:2^{T\hat R_i}], i\in \tilde{\NN}    \Big\}.
\end{align}

\emph{Encoding:} Let $\mathbf{m}=(m_1,m_2,\ldots,m_{B^{2}})$ be the message vector to be sent and let $m_b=1$ be the dummy message for any
\begin{align}
b\in  \cup_{w=1}^{B}[wB-L+1:wB] \bigcup  [(B-1)B +1:B^{2}]\label{E:dummy1}
\end{align}
and for any $b\leq 0$. As we will see, these dummy messages are inserted to ensure the start of block-by-block backward decoding. Due to these dummy messages,
the actually achievable rate becomes $\frac{(B-L)(B-1)}{B^2}R$, which, however, can be made arbitrarily close to $R$ by letting $B\to \infty$ for any $L$.

\begin{table}[htb!]\caption{Block-by-Block backward decoding for the single D-F relay node case}\label{tab}
\centering
\tiny
 \begin{tabular}{ccccccccc}
  \toprule
  Block &1&2&$\cdots$&$B-L$&$B-L+1$&$\cdots$&$B$&$\cdots$\\
  \midrule
  \vspace{4pt}
  $X_0$ & $\mathbf{x}_{0,1}(m_1|1)$ & $\mathbf{x}_{0,2}(m_2|1)$ & $\cdots$ & $\mathbf{x}_{0,B-L}(m_{B-L}|1)$ & $\mathbf{x}_{0,B-L+1}(1|1)$ & $\cdots$ & $\mathbf{x}_{0,B}(1|1)$&$\cdots$\\
  \vspace{4pt}
   $Y_1$ & $\emptyset$ & $\emptyset$ & $\cdots$ & $\emptyset$ & $\emptyset$ & $\cdots$ & $(m_1,m_2,\ldots,m_{B})$&$\cdots$\\
  \vspace{4pt}
  $X_1$ & $\mathbf{x}_{1,1}(1)$ & $\mathbf{x}_{1,2}(1)$ & $\cdots$ & $\mathbf{x}_{1,B-L}(1)$  & $\mathbf{x}_{1,B-L+1}(1)$  & $\cdots$ & $\mathbf{x}_{1,B}(1)$&$\cdots$\\
    \vspace{4pt}
  $Y_{\tilde{\NN}}$&$\hat{\mathbf{y}}_{\tilde{\NN},1}(l_{\tilde{\NN},1}|\mathbf{1})$&$\hat{\mathbf{y}}_{\tilde{\NN},2}(l_{\tilde{\NN},2}|l_{\tilde{\NN},1})$&$\cdots$&
  $\hat{\mathbf{y}}_{\tilde{\NN},B-L}(l_{\tilde{\NN},B-L}|l_{\tilde{\NN},B-L-1})$ &$\hat{\mathbf{y}}_{\tilde{\NN},B-L+1}(l_{\tilde{\NN},B-L+1}|l_{\tilde{\NN},B-L})$&$\cdots$& $\hat{\mathbf{y}}_{\tilde{\NN},B}(l_{\tilde{\NN},B}|l_{\tilde{\NN},B-1})$&$\cdots$\\
    \vspace{4pt}
  $X_{\tilde{\NN}}$ & $\mathbf{x}_{\tilde{\NN},1}(\mathbf{1})$ & $\mathbf{x}_{\tilde{\NN},2}(l_{\tilde{\NN},1})$ & $\cdots$ & $\mathbf{x}_{\tilde{\NN},B-L}(l_{\tilde{\NN},B-L-1})$
  & $\mathbf{x}_{\tilde{\NN},B-L+1}(l_{\tilde{\NN},B-L})$ & $\cdots$ & $\mathbf{x}_{\tilde{\NN},B}(l_{\tilde{\NN},B-1})$&$\cdots$ \\
  \vspace{4pt}
  $Y_{n+2}$ & $\emptyset$ & $\emptyset$ & $\cdots$ & $\emptyset$
  & $\emptyset$ & $\cdots$ & $\emptyset$&$\cdots$ \\
  \bottomrule
 \end{tabular}
   \vspace{4pt}

%  \vspace{4pt}
\vspace{10pt}
\begin{tabular}{cccccccc}
  \toprule
   Block & $B^2 - B +1$ & $\cdots$&  $B^2 - L$ & $B^2 - L+1$ & $\cdots$ & $B^2$\\
  \midrule
  \vspace{4pt}
  $X_0$ &$\mathbf{x}_{0,B^2 - B +1}(1|m_{B^2 - 2B +1})$ & $\cdots$&  $\mathbf{x}_{0,B^2 -L}(1|m_{B^2 -B-L})$ & $\mathbf{x}_{0,B^2 -L+1}(1|1)$ & $\cdots$ & $\mathbf{x}_{0,B^2}(1|1 )$ \\
  \vspace{4pt}
  $Y_1$ &  $\emptyset$ & $\cdots$& $\emptyset$ & $\emptyset$ & $\cdots$ & $(m_{B^2 - B +1}, \ldots, m_{B^2})$\\
  \vspace{4pt}
 $X_1$ & $\mathbf{x}_{1,B^2 - B +1}(m_{B^2 - 2B +1})$  & $\cdots$& $\mathbf{x}_{1,B^2 - L}(m_{B^2 -B-L})$  & $\mathbf{x}_{1,B^2 - L+1}(1)$  & $\cdots$ & $\mathbf{x}_{1,B^2}(1)$\\
    \vspace{4pt}
   $Y_{\tilde{\NN}}$& $\hat{\mathbf{y}}_{\tilde{\NN},B^2 - B +1}(l_{\tilde{\NN},B^2 - B +1}|l_{\tilde{\NN},B^2 - B })$ & $\cdots$&
   $\hat{\mathbf{y}}_{\tilde{\NN},B^2 - L}(l_{\tilde{\NN},B^2 - L}|l_{\tilde{\NN},B^2 - L-1})$ &
  $\hat{\mathbf{y}}_{\tilde{\NN},B^2 - L+1}(l_{\tilde{\NN},B^2 - L+1}|l_{\tilde{\NN},B^2 - L})$ &$\cdots$  & $\hat{\mathbf{y}}_{\tilde{\NN},B^2}(l_{\tilde{\NN},B^2}|l_{\tilde{\NN},B^2-1})$\\
    \vspace{4pt}
  $X_{\tilde{\NN}}$ &  $ \mathbf{x}_{\tilde{\NN},B^2 - B +1}(l_{\tilde{\NN},B^2 - B })$ & $\cdots$&
  $ \mathbf{x}_{\tilde{\NN},B^2 - L}(l_{\tilde{\NN},B^2 - L-1})$ &
  $ \mathbf{x}_{\tilde{\NN},B^2 - L+1}(l_{\tilde{\NN},B^2 - L})$ &$\cdots$  & $ \mathbf{x}_{\tilde{\NN},B^2}(l_{\tilde{\NN},B^2-1})$\\
   \vspace{4pt}
 $Y_{n+2}$ & $\emptyset$ & $\cdots$&
  $\emptyset$ & $\emptyset$  &$\cdots$  & $(m_1,m_2,\ldots,m_{B^2})$\\
  \bottomrule
 \end{tabular}
\end{table}

i) First consider the encoding process for nodes 0 and 1.
\begin{itemize}
  \item In block $b\in [1: B^{2}]$, the source node 0 transmits $\mathbf{x}_{0,b}(m_{b}|m_{b-B})$.
  \item At the end of block $vB,  v\in [1:B]$, the D-F relay node $1$ has decoded messages $$(m_{vB-B+1},m_{vB-B+2},\ldots,m_{vB})$$ using backward decoding (see the decoding part). In the next $B$ blocks, i.e., in block  $b\in [vB+1:(v+1)B]$, the relay node $1$ transmits $\mathbf{x}_{1,b}(m_{b-B})$, where $m_{b-B}$ for any $b\in [vB+1:(v+1)B])$ has been decoded by block $vB$.
\end{itemize}

ii) For any block $b\in [1: B^{2}]$, each relay node $i\in \tilde{\NN}$, upon receiving $\mathbf{y}_{i,b}$ at the end of block $b$, finds an index $l_{i,b}$ such that
$$(\mathbf{x}_{i,b}(l_{i,b-1}), \mathbf{y}_{i,b},\hat{\mathbf{y}}_{i,b}(l_{i,b}|l_{i,b-1}))\in A_{\epsilon}(X_i,Y_i,\hat{Y}_i),$$
where $l_{i,0}=1$ by convention. In block $b\in [1: B^{2}]$, the relay node $i\in \tilde{\NN}$ transmits $\mathbf{x}_{i,b}(l_{i,b-1})$.

\emph{Decoding:} We present the decoding process at the D-F relay node 1 and at the destination node $n+2$ separately.

i) At the end of block $b =vB,  v\in [1:B]$,  the D-F relay node 1 decodes messages $$(m_{b-B+1},m_{b-B+2},\ldots,m_b)$$ using block-by-block backward decoding. In fact, among these messages, $(m_{b-L+1},m_{b-L+2},\ldots,m_b)$ are dummy messages according to \dref{E:dummy1} and only $(m_{b-B+1},m_{b-B+2},\ldots,m_{b-L})$ need decoding.

\begin{itemize}
 \item a) Node $1$ first finds the unique $l_{\DD_1, b-L}=\{l_{i,b-L},i\in \DD_{1}\}$ such that there exists some $l_{\DD_1, b-L+1}^{b}$ satisfying that for any block $j=b-L+1,b-L+2,\ldots,b$,
\begin{align}
(&  \mathbf{X}_{0,j}(m_{j}|m_{j-B}), \mathbf{X}_{1,j}(m_{j-B}), \{(\mathbf{X}_{i,j}(l_{i,j-1}) , \hat{\mathbf{Y}}_{i,j}(l_{i,j}|l_{i,j-1})): i \in \DD_1 \}, \mathbf{Y}_{1,j}   )\in A_{\epsilon}(X_0, X_1, X_{\DD_1 }, \hat{Y}_{\DD_1 },Y_1). \label{E:chk1}
\end{align}

Note in \dref{E:chk1}, for any $j=b-L+1,b-L+2,\ldots,b$, $m_{j}$ and $m_{j-B}$  are both dummy messages according to \dref{E:dummy1}, and both $\mathbf{X}_{0,j}(m_{j}|m_{j-B})$ and $\mathbf{X}_{1,j}(m_{j-B})$ are known at node $1$. Then, it follows from the proof of \cite[Thm 2.7]{WuXieTIT} that $l_{\DD_1, b-L}$ can be decoded if
\begin{equation}\label{E:eq2singlerelay}
I(X_\SS;\hat Y_{\DD_{1} \setminus \SS},Y_{1}|X_{0},X_1, X_{\DD_{1} \setminus \SS})-I(Y_{\SS};\hat Y_{\SS}|X_{0},X_1,X_{\DD_{1}},Y_{1}, \hat Y_{\DD_{1} \setminus \SS})> 0,
\end{equation}
for any nonempty $\SS \subseteq \DD_{1}$.

 \item b) Backwardly and sequentially from block $j=b-L$ to $j=b-B+1$, node $1$ finds the unique pair $(m_{j},l_{\DD_1, j-1})$ satisfying \dref{E:chk1}, where $l_{\DD_1, j}$ has already been recovered due to the backward property of decoding, and $m_{j-B}$ has been decoded by block $b-B$.

$~~~$At each block $j=b-L,b-L-1,\ldots,b-B+1$, error occurs with $m_{j}$ if the true $m_{j}$ does not satisfy \dref{E:chk1} with any $l_{\DD_1, j-1}$, or a false $m_{j}$ satisfies \dref{E:chk1} with some $l_{\DD_1, j-1}$. According to the properties of typical sequences, the true $(m_{j},l_{\DD_1, j-1})$ satisfies \dref{E:chk1} with high probability.

\hspace{5mm}For a false $m_{j}$ and a $l_{\DD_1, j-1}$ with false $\{l_{i,j-1}, i\in \SS \}$ but true $\{l_{i,j-1}, i\in \DD_1 \setminus \SS \}$,
$\mathbf{X}_{0,j}(m_{j}|m_{j-B})$ is conditionally independent of $\{(\mathbf{X}_{i,j}(l_{i,j-1}) , \hat{\mathbf{Y}}_{i,j}(l_{i,j}|l_{i,j-1})): i \in \DD_1 \}$ and $\mathbf{Y}_{1,j}$  given
$\mathbf{X}_{1,j}(m_{j-B})$; and $\{(\mathbf{X}_{i,j}(l_{i,j-1}) , \hat{\mathbf{Y}}_{i,j}(l_{i,j}|l_{i,j-1})): i \in \SS \}$ are independent of
$\{(\mathbf{X}_{i,j}(l_{i,j-1}) , \hat{\mathbf{Y}}_{i,j}(l_{i,j}|l_{i,j-1})): i \in \DD_1 \setminus \SS \}$, $\mathbf{X}_{1,j}(m_{j-B})$ and $\mathbf{Y}_{1,j}$.

\hspace{5mm}Therefore, the probability that such false $(m_{j},l_{\DD_1, j-1})$ satisfies \dref{E:chk1} can be upper bounded by
\begin{align*}
&2^{T(H(X_0,X_1,X_{\DD_1},\hat Y_{\DD_1}, Y_{1} )+\epsilon)}2^{-T(H(X_{1},X_{\DD_1 \setminus \SS},\hat Y_{\DD_1 \setminus \SS}, Y_{1} )-\epsilon)}\\
\times &2^{-T(H(X_0| X_{1} )-\epsilon)}  2^{-T(H(X_{\SS})-\epsilon)}2^{-T(\sum_{i\in \SS}(H(\hat Y_{i}|X_i)-\epsilon))}.
\end{align*}
Since the number of such false $(m_{j},l_{\DD_1, j-1})$ is upper bounded by $2^{TR}\prod_{i\in \SS}2^{T(I(Y_i;\hat Y_i|X_i) +\epsilon)}$, with the union bound, it is easy to check that the probability of finding a false $m_{j}$ goes to zero as $T\to \infty$, if
\begin{align}\label{E:eq1singlerelay}
R< \min_{\SS \subseteq \DD_{1}}  I(X_0,X_\SS;\hat Y_{\DD_{1} \setminus \SS},Y_{1}|X_1, X_{\DD_{1} \setminus \SS} )-I(Y_{\SS};\hat Y_{\SS}|X_{0},X_1, X_{\DD_1},Y_{1}, \hat Y_{\DD_{1} \setminus \SS} ).
\end{align}
Then, based on the recovered $m_{j-B}$ and $l_{\DD_1, j}$, again from the proof of \cite[Thm 2.7]{WuXieTIT}, it follows that $l_{\DD_1, j-1}$ can be decoded if \dref{E:eq2singlerelay} holds.

\hspace{5mm}By a) and b) together, at the end of block $b =vB,  v\in [1:B]$, the D-F relay node 1 can decode messages $(m_{b-B+1},m_{b-B+2},\ldots,m_b)$ if both \dref{E:eq2singlerelay} and \dref{E:eq1singlerelay} hold.
\end{itemize}

ii) At the end of all $B^2$ block,  the destination node $n+2$ decodes messages $(m_{1},m_{2},\ldots,m_{B^2})$ using block-by-block backward decoding. Similarly, we only consider the decoding of $(m_{1},m_{2},\ldots,m_{B^2-B-L})$, since $(m_{B^2-B-L+1},m_{B^2-B-L+2},\ldots,m_{B^2})$ are all dummy messages according to \dref{E:dummy1}.
\begin{itemize}
 \item a) Node $n+2$ first finds the unique $l_{\DD_{n+2}, B^2-L}=\{l_{i,B^2-L},i\in \DD_{n+2}\}$ such that there exists some $l_{\DD_{n+2}, B^2-L+1}^{B^2}$ satisfying that for any block $j=B^2-L+1,B^2-L+2,\ldots,B^2$,
\begin{align}
&(\mathbf{X}_{0,j}(m_{j}|m_{j-B}), \mathbf{X}_{1,j}(m_{j-B}),  \{(\mathbf{X}_{i,j}(l_{i,j-1}) ,  \hat{\mathbf{Y}}_{i,j}(l_{i,j}|l_{i,j-1})): i \in \DD_{n+2} \}, \mathbf{Y}_{n+2,j} )
\nonumber \\
&\in A_{\epsilon}(X_0,X_1, X_{\DD_{n+2} }, \hat{Y}_{\DD_{n+2} },Y_{n+2}),\label{E:chk1'}
\end{align}
where, similarly, $m_{j}$ and $m_{j-B}$  are both dummy messages according to \dref{E:dummy1}, and $\mathbf{X}_{0,j}(m_{j}|m_{j-B})$ and $\mathbf{X}_{1,j}(m_{j-B})$ are both known at node $n+2$. Still, from the proof of \cite[Thm 2.7]{WuXieTIT}, $l_{\DD_{n+2}, B^2-L}$ can be decoded if
\begin{equation}\label{E:eq2'singlerelay}
I(X_\SS;\hat Y_{\DD_{n+2} \setminus \SS},Y_{n+2}|X_{0},X_1, X_{\DD_{n+2} \setminus \SS})-I(Y_{\SS};\hat Y_{\SS}|X_{0},X_1,X_{\DD_{n+2}},Y_{n+2}, \hat Y_{\DD_{n+2} \setminus \SS})> 0,
\end{equation}
for any nonempty $\SS \subseteq \DD_{n+2}$.

 \item b) Backwardly and sequentially from block $j=B^2-L$ to $j=1$, node $n+2$ finds the unique pair $(m_{j-B},l_{\DD_n+2, j-1})$ satisfying \dref{E:chk1'}, where $l_{\DD_{n+2}, j}$ has already been recovered due to the backward property of decoding, and $m_{j}$ either is a dummy message (for $j=B^2-L,B^2-L-1,\ldots,B^2-B-L+1$) or has been decoded due to the backward property of decoding (for $j=B^2-B-L,B^2-B-L-1,\ldots,1$).

\hspace{5mm}At each block $j=B^2-L,B^2-L-1,\ldots,1$, error occurs with $m_{j-B}$ if the true $m_{j-B}$ does not satisfy \dref{E:chk1'} with any $l_{\DD_{n+2}, j-1}$, or a false $m_{j-B}$ satisfies \dref{E:chk1'} with some $l_{\DD_{n+2}, j-1}$. According to the properties of typical sequences, the true $(m_{j-B},l_{\DD_{n+2}, j-1})$ satisfies \dref{E:chk1'} with high probability.

\hspace{5mm}For a false $m_{j-B}$ and a $l_{\DD_{n+2}, j-1}$ with false $\{l_{i,j-1}, i\in \SS \}$ but true $\{l_{i,j-1}, i\in \DD_{n+2} \setminus \SS \}$,
$\mathbf{X}_{0,j}(m_{j}|m_{j-B})$ and $\mathbf{X}_{1,j}(m_{j-B})$ are independent of $\{(\mathbf{X}_{i,j}(l_{i,j-1}) , \hat{\mathbf{Y}}_{i,j}(l_{i,j}|l_{i,j-1})): i \in \DD_{n+2} \}$ and $\mathbf{Y}_{n+2,j}$; and $\{(\mathbf{X}_{i,j}(l_{i,j-1}) , \hat{\mathbf{Y}}_{i,j}(l_{i,j}|l_{i,j-1})): i \in \SS \}$ are independent of
$\{(\mathbf{X}_{i,j}(l_{i,j-1}) , \hat{\mathbf{Y}}_{i,j}(l_{i,j}|l_{i,j-1})): i \in \DD_{n+2} \setminus \SS \}$ and $\mathbf{Y}_{n+2,j}$.

\hspace{5mm}Therefore, the probability that such false $(m_{j},l_{\DD_{n+2}, j-1})$ satisfies \dref{E:chk1'} can be upper bounded by
\begin{align*}
&2^{T(H(X_0,X_1,X_{\DD_{n+2}},\hat Y_{\DD_{n+2}}, Y_{n+2} )+\epsilon)}2^{-T(H(X_{\DD_{n+2} \setminus \SS},\hat Y_{\DD_{n+2} \setminus \SS}, Y_{n+2} )-\epsilon)}\\
\times &2^{-T(H(X_0, X_{1} )-\epsilon)}  2^{-T(H(X_{\SS})-\epsilon)}2^{-T(\sum_{i\in \SS}(H(\hat Y_{i}|X_i)-\epsilon))}.
\end{align*}
Since the number of such false $(m_{j},l_{\DD_{n+2}, j-1})$ is upper bounded by $2^{TR}\prod_{i\in \SS}2^{T(I(Y_i;\hat Y_i|X_i) +\epsilon)}$, with the union bound, it is easy to check that the probability of finding a false $m_{j}$ goes to zero as $T\to \infty$, if
\begin{align}\label{E:eq1'singlerelay}
R< \min_{\SS \subseteq \DD_{1}}  I(X_0,X_1,X_\SS;\hat Y_{\DD_{n+2} \setminus \SS},Y_{n+2}| X_{\DD_{n+2} \setminus \SS} )-I(Y_{\SS};\hat Y_{\SS}|X_{0},X_1, X_{\DD_{n+2}},Y_{n+2}, \hat Y_{\DD_{n+2} \setminus \SS} ).
\end{align}
Similarly, based on the recovered $m_{j}$ and $l_{\DD_{n+2}, j}$, $l_{\DD_{n+2}, j-1}$ can be decoded if \dref{E:eq2'singlerelay} holds.

\hspace{5mm}By a) and b) together, at the end of all $B^2$ block, the destination node $n+2$ can decode messages $(m_1,m_{2},\ldots,m_{B^2})$ if both \dref{E:eq2'singlerelay} and \dref{E:eq1'singlerelay} hold.
\end{itemize}

Combining i) and ii), and using the standard technique of time sharing, we conclude that the rate described in \dref{E:simplified}-\dref{E:sim2} is achievable.

\subsection{Multiple D-F relay nodes ($M\geq 2$)}

When there are multiple D-F relay nodes, i.e., $M\geq 2$,  a total of $B^{M+1}$ blocks will be used. The detailed codebook generation and encoding/decoding process are as follows.

\emph{Codebook Generation:} Fix $p(x_0)p(x_{ \MM}|x_0)\prod_{i\in \NN \setminus \MM}p(x_i)p(\hat y_i|y_i,x_i)$. We randomly and independently generate a codebook for each block.

i) First consider the codebook generation for nodes $\pi(1:M+1)$.
\begin{itemize}
  \item For each block $b\in [1: B^{M+1}]$, backwardly and sequentially for each relay node $\pi(k), k=M+1,M,\ldots,2$, randomly generate
$2^{TR}$ conditionally independent sequences $$\mathbf{x}_{\pi(k),b}(m_{b-B^{k-1}}|m_{b-B^{k}},\ldots,m_{b-B^{M}}),$$ where $m_{b-B^{k-1}},m_{b-B^{k}},\ldots,m_{b-B^{M}}\in [1:2^{TR}]$;
  \item For each block $b\in [1: B^{M+1}]$ and node $\pi(1)$, i.e., the source node 0, randomly generate $2^{TR}$ conditionally independent sequences $\mathbf{x}_{0,b}(m_{b}|m_{b-B},\ldots,m_{b-B^{M}})$, where $$m_{b},m_{b-B},\ldots,m_{b-B^{M}}\in [1:2^{TR}].$$
\end{itemize}

ii) The codebook generation for the nodes in $\NN \setminus \MM$ is the same as that in the case of $M=1$. For each block $b\in [1: B^{M+1}]$ and each relay node $i\in \NN \setminus \MM$, randomly and independently generate $2^{T\hat R_i}$ sequences $\mathbf{x}_{i,b}(l_{i,b-1})$, $l_{i,b-1}\in [1:2^{T\hat R_i}]$, where $\hat R_i=I(Y_i;\hat Y_i|X_i)+\epsilon$;
for each relay node $i\in \NN \setminus \MM$ and each $\mathbf{x}_{i,b}(l_{i,b-1})$, $l_{i,b-1}\in [1:2^{T\hat R_i}]$, randomly and conditionally independently generate $2^{T\hat R_i}$ sequences $ \hat{\mathbf{y}}_{i,b}(l_{i,b}|l_{i,b-1})$, $l_{i,b}\in [1:2^{T\hat R_i}]$.

The combination of i) and ii) defines the codebook for any block $b\in [1: B^{M+1}]$,
\begin{align}
\mathcal{C}_b=\Big\{&\mathbf{x}_{\pi(k),b}(m_{b-B^{k-1}}|m_{b-B^{k}},\ldots,m_{b-B^{M}}): m_{b-B^{k-1}},m_{b-B^{k}},\ldots,m_{b-B^{M}}\in [1:2^{TR}], k=M+1,M,\ldots,2;\nonumber \\
&\mathbf{x}_{0,b}(m_{b}|m_{b-B},\ldots,m_{b-B^{M}}):m_{b},m_{b-B},\ldots,m_{b-B^{M}}\in [1:2^{TR}];\nonumber \\
 &\mathbf{x}_{i,b}(l_{i,b-1}) , \hat{\mathbf{y}}_{i,b}(l_{i,b}|l_{i,b-1}):  l_{i,b},l_{i,b-1} \in  [1:2^{T\hat R_i}], i\in \NN \setminus \MM       \Big\}.
\end{align}

\emph{Encoding:} Let $\mathbf{m}=(m_1,m_2,\ldots,m_{B^{M+1}})$ be the message vector to be sent and let $m_b=1$ be the dummy message for any
\begin{align}
b\in  \cup_{w=1}^{B^M}[wB-L+1:wB] \bigcup   \cup_{u=1}^{M} \cup_{v=1}^{B^{M-u}} [v(B-1)B^u +1:vB^{u+1}],\label{E:dummy}
\end{align}
and for any $b\leq 0$. Now, the actually achievable rate is $\frac{B-L}{B} (\frac{B-1}{B})^M R$ due to the dummy messages, which can still be made arbitrarily close to $R$ by letting $B\to \infty$ for any $L$ and $M$.

i) We still first consider the encoding process for nodes $\pi(1:M+1)$.
\begin{itemize}
  \item In block $b\in [1: B^{M+1}]$, node $\pi(1)$, i.e., the source node 0, transmits $\mathbf{x}_{0,b}(m_{b}|m_{b-B},\ldots,m_{b-B^{M}})$.
  \item By the end of block $vB^{k-1},  v\in [1:B^{M+1}/B^{k-1}]$, the D-F relay node $\pi(k),k=2,\ldots,M+1$, has decoded messages $(m_{1},m_{2},\ldots,m_{vB^{k-1}})$ using backward decoding (see the decoding part). In the next $B^{k-1}$ blocks, i.e., in block  $b\in [vB^{k-1}+1:(v+1)B^{k-1}]$, node $\pi(k),k=2,\ldots,M+1$, transmits $\mathbf{x}_{\pi(k),b}(m_{b-B^{k-1}}|m_{b-B^{k}},\ldots,m_{b-B^{M}})$, where $$(m_{b-B^{k-1}},m_{b-B^{k}},\ldots,m_{b-B^{M}}), b\in [vB^{k-1}+1:(v+1)B^{k-1}]$$ have all been decoded by block $vB^{k-1}$.
\end{itemize}

ii) The encoding process for the nodes in $\NN \setminus \MM$ is still the same as that in the case of $M=1$. For any block $b\in [1: B^{M+1}]$, each relay node $i\in \NN \setminus \MM$, upon receiving $\mathbf{y}_{i,b}$ at the end of block $b$, finds an index $l_{i,b}$ such that
$$(\mathbf{x}_{i,b}(l_{i,b-1}), \mathbf{y}_{i,b},\hat{\mathbf{y}}_{i,b}(l_{i,b}|l_{i,b-1}))\in A_{\epsilon}(X_i,Y_i,\hat{Y}_i),$$
where $l_{i,0}=1$ by convention. In block $b\in [1: B^{M+1}]$, the relay node $i\in \NN \setminus \MM$ transmits $\mathbf{x}_{i,b}(l_{i,b-1})$.

\emph{Decoding:} At the end of block $b =vB^{k-1},  v\in [1:B^{M+1}/B^{k-1}]$, the node $\pi(k),k=2,\ldots,M+2$, decodes messages $(m_{b-B^{k-1}+1},\ldots,m_b)$ using block-by-block backward decoding as follows.

i) The node $\pi(k),k=2,\ldots,M+2$,  first finds the unique $l_{\DD_k, b-L}=\{l_{i,b-L},i\in \DD_{k}\}$ such that there exists some $l_{\DD_k, b-L+1}^{b}$ satisfying that for any block $j=b-L+1,b-L+2,\ldots,b$,
\begin{align}
(&  \mathbf{X}_{0,j}(m_{j}|m_{j-B},\ldots,m_{j-B^{M}}), \nonumber\\
&\{\mathbf{X}_{\pi(s),j}(m_{j-B^{s-1}}|m_{j-B^{s}},\ldots,m_{j-B^{M}}),s=2,\ldots,k-1,k,k+1,\ldots,M+1\}, \nonumber \\
& \{(\mathbf{X}_{i,j}(l_{i,j-1}) , \hat{\mathbf{Y}}_{i,j}(l_{i,j}|l_{i,j-1})): i \in \DD_k \}, \mathbf{Y}_{\pi(k),j}   )\in A_{\epsilon}(X_0,X_\MM,X_{\DD_k }, \hat{Y}_{\DD_k },Y_{\pi(k)}). \label{E:chk}
\end{align}

Note in \dref{E:chk}, $(m_{j},m_{j-B},\ldots,m_{j-B^{M}}), j=b-L+1,b-L+2,\ldots,b$ are all dummy messages according to \dref{E:dummy}, and thus $\mathbf{X}_{\pi(s),j},s=1,\ldots,M+1$ are all known at node $\pi(k)$. Then, it follows from the proof of \cite[Thm 2.7]{WuXieTIT} that $l_{\DD_k, b-L}$ can be decoded if
\begin{equation}
I(X_\SS;\hat Y_{\DD_{k} \setminus \SS},Y_{\pi(k)}|X_{\pi(1:M+1)},X_{\DD_{k} \setminus \SS})-I(Y_{\SS};\hat Y_{\SS}|X_{\pi(1:M+1)},X_{\DD_{k}},Y_{\pi(k)}, \hat Y_{\DD_{k} \setminus \SS})> 0, \label{E:constraint2''}
\end{equation}
for any nonempty $\SS \subseteq \DD_{k}$.

ii) Backwardly and sequentially from block $j=b-L$ to $j=b-B^{k-1}+1$, node $\pi(k),k=2,3,\ldots,M+2$, jointly decodes the message transmitted by its immediate upstream node $\pi(k-1)$, and the compressions of the C-F relay nodes. Specifically, node $\pi(k),k=2,3,\ldots,M+2$ finds the unique pair $(m_{j-B^{k-2}},l_{\DD_k, j-1})$ satisfying \dref{E:chk}; node $\pi(k),k=2$, finds the unique pair $(m_{j},l_{\DD_k, j-1})$ satisfying \dref{E:chk}. Note here the exception for node $\pi(2)$ arises  because the source node $\pi(1)$ transmits $m_j$ rather than $m_{j-1}$ in block $j$, but the ideas of the decoding processes at all $\pi(k),k=2,3,\ldots,M+2$, are exactly the same. Thus, below, we only present the decoding at node $\pi(k),k=3,4,\ldots,M+2$, while the decoding at node $\pi(2)$ can be easily obtained by analogy. The same consideration also applies to the proof in \ref{subsection}.

In  \dref{E:chk}, $l_{\DD_k, j}$ has already been recovered due to the backward property of decoding, and among the messages $(m_{j},m_{j-B},\ldots,m_{j-B^{M}})$, only $m_{j-B^{k-2}}$ is the unknown message at node $\pi(k)$ that needs to be decoded in block $j$. In fact, $(m_{j-B^{k-1}},\ldots, m_{j-B^{M}})$ have been decoded by block $b-B^{k-1}$, while $(m_{j},\ldots, m_{j-B^{k-3}})$ either are dummy messages according to \dref{E:dummy} (for block $j=b-L,b-L-1 \ldots, b-B^{k-2}-L+1$) or have been decoded due to the backward property of decoding (for block $j=b-B^{k-2}-L,b-B^{k-2}-L-1,\ldots,b-B^{k-1}+1$).

At each block $j=b-L,b-L-1,\ldots,b-B^{k-1}+1$, error occurs with $m_{j-B^{k-2}}$ if the true $m_{j-B^{k-2}}$ does not satisfy \dref{E:chk} with any $l_{\DD_k, j-1}$, or a false $m_{j-B^{k-2}}$ satisfies \dref{E:chk} with some $l_{\DD_k, j-1}$. According to the properties of typical sequences, the true $(m_{j-B^{k-2}},l_{\DD_k, j-1})$ satisfies \dref{E:chk} with high probability.

For a false $m_{j-B^{k-2}}$ and a $l_{\DD_k, j-1}$ with false $\{l_{i,b-1}, i\in \SS \}$ but true $\{l_{i,b-1}, i\in \DD_k \setminus \SS \}$,
$$\{  \mathbf{X}_{\pi(1),j}(m_{j}|m_{j-B},\ldots,m_{j-B^{M}}),   \mathbf{X}_{\pi(s),j}(m_{j-B^{s-1}}|m_{j-B^{s}},\ldots,m_{j-B^{M}}),s=2,\ldots,k-1\}$$ are conditionally independent of $\{(\mathbf{X}_{i,j}(l_{i,j-1}) , \hat{\mathbf{Y}}_{i,j}(l_{i,j}|l_{i,j-1})): i \in \DD_k \}$ and $\mathbf{Y}_{\pi(k),j}$  given
$$\{\mathbf{X}_{\pi(s),j}(m_{j-B^{s-1}}|m_{j-B^{s}},\ldots,m_{j-B^{M}}),s=k,\ldots,M+1\};$$
and $\{(\mathbf{X}_{i,j}(l_{i,j-1}) , \hat{\mathbf{Y}}_{i,j}(l_{i,j}|l_{i,j-1})): i \in \SS \}$ are independent of
$$\{(\mathbf{X}_{i,j}(l_{i,j-1}) , \hat{\mathbf{Y}}_{i,j}(l_{i,j}|l_{i,j-1})): i \in \DD_k \setminus \SS \}, \{\mathbf{X}_{\pi(s),j}(m_{j-B^{s-1}}|m_{j-B^{s}},\ldots,m_{j-B^{M}}),s\in [k:M+1]\}, \mathbf{Y}_{\pi(k),j}.$$

Therefore, the probability that such false $(m_{j-B^{k-2}},l_{\DD_k, j-1})$ satisfies \dref{E:chk} can be upper bounded by
\begin{align*}
&2^{T(H(X_{\pi(1:M+1)},X_{\DD_k},\hat Y_{\DD_k}, Y_{\pi(k)} )+\epsilon)}2^{-T(H(X_{\pi(k:M+1)},X_{\DD_k \setminus \SS},\hat Y_{\DD_k \setminus \SS}, Y_{\pi(k)} )-\epsilon)}\\
\times &2^{-T(H(X_{\pi(1:k-1)}| X_{\pi(k:M+1)} )-\epsilon)}  2^{-T(H(X_{\SS})-\epsilon)}2^{-T(\sum_{i\in \SS}(H(\hat Y_{i}|X_i)-\epsilon))}.
\end{align*}
Since the number of such false $(m_{j-B^{k-2}},l_{\DD_k, j-1})$ is upper bounded by $2^{TR}\prod_{i\in \SS}2^{T(I(Y_i;\hat Y_i|X_i) +\epsilon)}$, with the union bound, it is easy to check that the probability of finding a false $m_{j-B^{k-2}}$ goes to zero as $T\to \infty$, if
\begin{equation}
R<\min_{\SS \subseteq \DD_{k}}I(X_{\pi(1:k-1)},X_\SS;\hat Y_{\DD_{k} \setminus \SS},Y_{\pi(k)}|X_{\DD_{k} \setminus \SS},X_{\pi(k:M+1)})-I(Y_{\SS};\hat Y_{\SS}|X_{\pi(1:M+1)},X_{\DD_{k}},Y_{\pi(k)}, \hat Y_{\DD_{k} \setminus \SS}). \label{E:constraint1''}
\end{equation}
Then, based on the recovered $m_{j-B^{k-2}}$ and $l_{\DD_k, j}$, from the proof  of \cite[Thm 2.7]{WuXieTIT}, it follows that $l_{\DD_k, j-1}$ can be decoded if \dref{E:constraint2''} holds.

Combining i) and ii), using the technique of time sharing, we obtain the achievable rate \dref{E:constraint1}-\dref{E:constraint2}.

\section{Unified Relay Framework With $B$-Blocks-By-$B$-Blocks Backward Decoding}\label{S:Bblocks}
Under the unified relay framework using nested blocks and backward decoding, we can also consider combining the noisy network coding scheme \cite{KimElGamal} with the multi-level D-F scheme. However, since noisy network coding uses repetitive encoding/all blocks united decoding, to make it fit into our framework, a modification is needed. Specifically, assume some fixed $\MM \subseteq \NN$ with $|\MM|=M$ and $\pi(\{0,\MM,n+1\})$, and a total of $B^{M+1}$ blocks are used. The source can still repetitively encode intra-$B$-blocks as in \cite{KimElGamal}, but inter-$B$-blocks, the source has to cumulatively encode to allow for the operation of D-F strategy; Correspondingly, both the D-F relay nodes and the destination will perform $B$-Blocks-By-$B$-Blocks backward decoding, which is essentially a combination of backward decoding and $B$ blocks united decoding. Same as in Section \ref{S:byb}, the backward decoding at node $\pi(k),k=2,3,\ldots,M+2$, will happen at the end of every $B^{k-1}$ blocks, i.e., at the end of block $b =vB^{k-1},  v\in [1:B^{M+1}/B^{k-1}]$, and  both the D-F relay nodes and the destination node perform compression-message joint decoding. Below, we still first consider the case of single D-F relay node ($M=1$) to illustrate the main idea, and then extend it to the general case of multiple D-F relay nodes ($M\geq 2$).

\subsection{Single D-F relay node ($M=1$)}\label{S:BbyBsingle}
Still assume that only node 1 is the D-F relay node, and all other relay nodes are the C-F relay nodes, and let $\tilde{\NN}:=\NN \setminus \{1\}$. Specializing Theorem  \ref{main2} to this case, we have that a rate $R$ is achievable, if there exists  some $$p(q)p(x_0|q)p(x_1|x_0,q)\prod_{i\in \tilde{\NN}}p(x_i|q)p(\hat y_i|y_i,x_i,q),$$
such that
\begin{equation}
R<\min \left\{
\begin{split}
& \max_{\TT_1 \subseteq \tilde{\NN}} \min_{\SS \subseteq \TT_{1}}  I(X_0,X_\SS;\hat Y_{\TT_{1} \setminus \SS},Y_{1}|X_1, X_{\TT_{1} \setminus \SS},Q)-I(Y_{\SS};\hat Y_{\SS}|X_{0},X_1, X_{\TT_1},Y_{1}, \hat Y_{\TT_{1} \setminus \SS},Q), \\
&\max_{\TT_{n+2} \subseteq \tilde{\NN}} \min_{\SS \subseteq \TT_{n+2}}  I(X_0,X_1,X_\SS;\hat Y_{\TT_{n+2} \setminus \SS},Y_{n+2}|X_{\TT_{n+2} \setminus \SS},Q)-I(Y_{\SS};\hat Y_{\SS}|X_{0},X_1, X_{\TT_{n+2}},Y_{n+2}, \hat Y_{\TT_{n+2} \setminus \SS},Q).
\end{split}
\right\}\label{E:simplified'}
\end{equation}

Still, a total of $B^{2}$ blocks will be used. The detailed codebook generation and encoding/decoding process are as follows, which can be understood with the help of Table \ref{tab2}.

\begin{table}[htb!]\caption{$B$-Blocks-by-$B$-Blocks backward decoding for the single D-F relay node case}\label{tab2}
\centering
\scriptsize
 \begin{tabular}{ccccccc}
  \toprule
  Block &1&2&$\cdots$&$B-1$&$B$&$\cdots$\\
  \midrule
  \vspace{4pt}
  $X_0$ & $\mathbf{x}_{0,1}(m_1|1)$ & $\mathbf{x}_{0,2}(m_1|1)$ & $\cdots$ & $\mathbf{x}_{0,B-1}(m_{1}|1)$ & $\mathbf{x}_{0,B}(m_{1}|1)$&$\cdots$ \\
  \vspace{4pt}
   $Y_1$ & $\emptyset$ & $\emptyset$ & $\cdots$ & $\emptyset$ & $m_1$&$\cdots$\\
  \vspace{4pt}
  $X_1$ & $\mathbf{x}_{1,1}(1)$ & $\mathbf{x}_{1,2}(1)$ & $\cdots$ & $\mathbf{x}_{1,B-1}(1)$  & $\mathbf{x}_{1,B}(1)$ &$\cdots$ \\
    \vspace{4pt}
  $Y_{\tilde{\NN}}$&$\hat{\mathbf{y}}_{\tilde{\NN},1}(l_{\tilde{\NN},1}|\mathbf{1})$&$\hat{\mathbf{y}}_{\tilde{\NN},2}(l_{\tilde{\NN},2}|l_{\tilde{\NN},1})$&$\cdots$&
  $\hat{\mathbf{y}}_{\tilde{\NN},B-1}(l_{\tilde{\NN},B-1}|l_{\tilde{\NN},B-2})$ &$\hat{\mathbf{y}}_{\tilde{\NN},B }(l_{\tilde{\NN},B}|l_{\tilde{\NN},B-1})$ &$\cdots$\\
    \vspace{4pt}
  $X_{\tilde{\NN}}$ & $\mathbf{x}_{\tilde{\NN},1}(\mathbf{1})$ & $\mathbf{x}_{\tilde{\NN},2}(l_{\tilde{\NN},1})$ & $\cdots$ & $\mathbf{x}_{\tilde{\NN},B-1}(l_{\tilde{\NN},B-2})$
  & $\mathbf{x}_{\tilde{\NN},B }(l_{\tilde{\NN},B-1})$  &$\cdots$\\
  \vspace{4pt}
  $Y_{n+2}$ & $\emptyset$ & $\emptyset$ & $\cdots$ & $\emptyset$
  & $\emptyset$ &$\cdots$ \\
  \bottomrule
 \end{tabular}
   \vspace{4pt}
%$\vdots~~~~$
  \vspace{4pt}
 \begin{tabular}{cccccc}
  \toprule
  Block &$B^2-B+1$&$B^2-B+2$&$\cdots$&$B^2-1$&$B^2$\\
  \midrule
  \vspace{4pt}
  $X_0$ & $\mathbf{x}_{0,B^2-B+1}(1|m_{B-1})$ & $\mathbf{x}_{0,B^2-B+2}(1|m_{B-1})$ & $\cdots$ & $\mathbf{x}_{0,B^2-1}(1|m_{B-1})$ & $\mathbf{x}_{0,B^2}(1|m_{B-1})$ \\
  \vspace{4pt}
   $Y_1$ & $\emptyset$ & $\emptyset$ & $\cdots$ & $\emptyset$ & $m_B$\\
  \vspace{4pt}
  $X_1$ & $\mathbf{x}_{1,B^2-B+1}(m_{B-1})$ & $\mathbf{x}_{1,B^2-B+2}(m_{B-1})$ & $\cdots$ & $\mathbf{x}_{1,B^2-1}(m_{B-1})$  & $\mathbf{x}_{1,B^2}(m_{B-1})$  \\
    \vspace{4pt}
  $Y_{\tilde{\NN}}$&$\hat{\mathbf{y}}_{\tilde{\NN},B^2-B+1}(l_{\tilde{\NN},B^2-B+1}|l_{\tilde{\NN},B^2-B})$&$\hat{\mathbf{y}}_{\tilde{\NN},B^2-B+2}(l_{\tilde{\NN},B^2-B+2}|l_{\tilde{\NN},B^2-B+1})$&$\cdots$&
  $\hat{\mathbf{y}}_{\tilde{\NN},B^2-1}(l_{\tilde{\NN},B^2-1}|l_{\tilde{\NN},B^2-2})$ &$\hat{\mathbf{y}}_{\tilde{\NN},B^2 }(l_{\tilde{\NN},B^2}|l_{\tilde{\NN},B^2-1})$ \\
    \vspace{4pt}
  $X_{\tilde{\NN}}$ & $\mathbf{x}_{\tilde{\NN},B^2-B+1}(l_{\tilde{\NN},B^2-B})$ & $\mathbf{x}_{\tilde{\NN},B^2-B+2}(l_{\tilde{\NN},B^2-B+1})$ & $\cdots$ & $\mathbf{x}_{\tilde{\NN},B^2-1}(l_{\tilde{\NN},B^2-2})$
  & $\mathbf{x}_{\tilde{\NN},B^2 }(l_{\tilde{\NN},B^2-1})$  \\
  \vspace{4pt}
  $Y_{n+2}$ & $\emptyset$ & $\emptyset$ & $\cdots$ & $\emptyset$ & $(m_1,m_2,\ldots,m_B)$
   \\
  \bottomrule
 \end{tabular}
\end{table}

\emph{Codebook Generation:}  Fix $p(x_0)p(x_1|x_0)\prod_{i\in \tilde{\NN}}p(x_i)p(\hat y_i|y_i,x_i)$. We randomly and independently generate a codebook for each block.

i) First consider the codebook generation for the source node 0 and the D-F relay node 1. Denote $f(b):=\lceil \frac{b}{B}\rceil$, i.e., the smallest integer greater than or equal to $\frac{b}{B}$.
For each block $b\in [1: B^{2}]$,  randomly generate $2^{TBR}$ independent sequences $\mathbf{x}_{1,b}(m_{f(b-B)})$ for node 1, and randomly generate $2^{TBR}$ conditionally independent sequences $\mathbf{x}_{0,b}(m_{f(b)}|m_{f(b-B)})$ for node 0, where $m_{f(b)},m_{f(b-B)}\in [1:2^{TBR}]$.

ii) The codebook generation for the C-F relay nodes is exactly the same as that in Section \ref{S:byb}.
For each block $b\in [1: B^{2}]$ and each relay node $i\in \tilde{\NN}$, randomly and independently generate $2^{T\hat R_i}$ sequences $\mathbf{x}_{i,b}(l_{i,b-1})$, $l_{i,b-1}\in [1:2^{T\hat R_i}]$, where $\hat R_i=I(Y_i;\hat Y_i|X_i)+\epsilon$;
for each relay node $i\in \tilde{\NN}$ and each $\mathbf{x}_{i,b}(l_{i,b-1})$, $l_{i,b-1}\in [1:2^{T\hat R_i}]$, randomly and conditionally independently generate $2^{T\hat R_i}$ sequences $ \hat{\mathbf{y}}_{i,b}(l_{i,b}|l_{i,b-1})$, $l_{i,b}\in [1:2^{T\hat R_i}]$.

The combination of i) and ii) defines the codebook for any block $b\in [1: B^{2}]$,
\begin{align}
\mathcal{C}_b=\Big\{&\mathbf{x}_{1,b}(m_{f(b-B)}), \mathbf{x}_{0,b}(m_{f(b)}|m_{f(b-B)}): m_{f(b)},m_{f(b-B)}\in [1:2^{TBR}];  \nonumber \\
 &\mathbf{x}_{i,b}(l_{i,b-1}) , \hat{\mathbf{y}}_{i,b}(l_{i,b}|l_{i,b-1}):  l_{i,b},l_{i,b-1} \in  [1:2^{T\hat R_i}], i\in \tilde{\NN}  \Big\}.
\end{align}

\emph{Encoding:} Let the message vector to be sent be $$\mathbf{m}=(\underbrace{m_1,m_1,\ldots,m_1}_{B},\underbrace{m_2,m_2,\ldots,m_2}_{B},\ldots,\underbrace{m_{B},m_{B},\ldots,m_{B}}_{B}).$$ Let $m_B=1$ be the dummy message,  i.e., $m_{f(b)}=1$ for any
\begin{align}
b\in  [ (B-1)B  +1:B^2],\label{E:dummy'}
\end{align}
and for any $b\leq 0$. The actually achievable rate is $  \frac{B-1}{B} R$ due to the dummy messages, which, however, can be made arbitrarily close to $R$ by letting $B\to \infty.$

i) First consider the encoding process for nodes 0 and 1.
\begin{itemize}
  \item In block $b\in [1: B^{2}]$, the source node 0 transmits $\mathbf{x}_{0,f(b)}(m_{f(b)}|m_{f(b-B)})$.
  \item At the end of block $vB,  v\in [1:B]$, the D-F relay node $1$ has decoded messages $m_{v}$ using $B$ blocks united decoding (see the decoding part). In the next $B$ blocks, i.e., in block  $b\in [vB+1:(v+1)B]$, the relay node $1$ transmits $\mathbf{x}_{1,b}(m_{f(b-B)})$, where $m_{f(b-B)}$  for any $b\in [vB+1:(v+1)B]$ is corresponding to $m_{v})$ that has been decoded by block $vB$.
\end{itemize}

ii) For any block $b\in [1: B^{2}]$, each relay node $i\in \tilde{\NN}$, upon receiving $\mathbf{y}_{i,b}$ at the end of block $b$, finds an index $l_{i,b}$ such that
$$(\mathbf{x}_{i,b}(l_{i,b-1}), \mathbf{y}_{i,b},\hat{\mathbf{y}}_{i,b}(l_{i,b}|l_{i,b-1}))\in A_{\epsilon}(X_i,Y_i,\hat{Y}_i),$$
where $l_{i,0}=1$ by convention. In block $b\in [1: B^{2}]$, the relay node $i\in \tilde{\NN}$ transmits $\mathbf{x}_{i,b}(l_{i,b-1})$.

\emph{Decoding:} We present the decoding process at the D-F relay node 1 and at the destination node $n+2$ separately.

i) At the end of block $b =vB,  v\in [1:B]$,  the D-F relay node 1 decodes messages $m_v$ using $B$ blocks united decoding, i.e., it finds the unique $m_v$, such that there exists some $l_{\tilde{\NN}, (v-1)B+1}^{vB}$ satisfying that for any block $j=(v-1)B+1,(v-1)B+2,\ldots,vB$,
\begin{align}
(&  \mathbf{X}_{0,j}(m_{f(j)}|m_{f(j-B)}),  \mathbf{X}_{1,j}(m_{f(j-B)}) \nonumber \\
& \{(\mathbf{X}_{i,j}(l_{i,j-1}) , \hat{\mathbf{Y}}_{i,j}(l_{i,j}|l_{i,j-1})): i \in \tilde{\NN} \}, \mathbf{Y}_{1,j}   )\in A_{\epsilon}(X_0,X_1,X_{\tilde{\NN} }, \hat{Y}_{\tilde{\NN} },Y_{1}), \label{E:chk'}
\end{align}
where $m_{f(j-B)}$ is corresponding to $m_{v-1}$ and has been decoded by the end of block $(v-1)B$, and $m_{f(j)}$ is corresponding to $m_v$. From \cite[Thm 1]{KimElGamal} and its proof (see also \cite[Thm 2.4]{WuXieTIT}), we have that $m_v$ can be decoded if
\begin{align}
R<\min_{\SS \subseteq \tilde{\NN}}I(X_{0},X_\SS;\hat Y_{\tilde{\NN} \setminus \SS},Y_{1}|X_{\tilde{\NN} \setminus \SS},X_{1})-I(Y_{\SS};\hat Y_{\SS}|X_0,X_1,X_{\tilde{\NN}},Y_{1}, \hat Y_{\tilde{\NN} \setminus \SS}). \label{tobeimproved1}
\end{align}
Note, \dref{tobeimproved1} can be improved by considering only a subset $\TT_1 \subseteq \tilde{\NN}$ for the decoding while treating the inputs of other C-F relay nodes as purely noise, leading to following more general rate constraint:
\begin{align}
R<\max_{\TT_1 \subseteq \tilde{\NN}}\min_{\SS \subseteq \TT_1}I(X_{0},X_\SS; \hat Y_{\TT_1 \setminus \SS},Y_{1}|X_{\TT_1 \setminus \SS},X_{1})-I(Y_{\SS};\hat Y_{\SS}|X_0,X_1,X_{\TT_1},Y_{1}, \hat Y_{\TT_1 \setminus \SS}). \label{hasimproved1}
\end{align}

ii) At the end of all $B^2$ block,  the destination node decodes all messages $(m_{1},m_{2},\ldots,m_{B})$ using $B$-Blocks-By-$B$-Blocks backward decoding. In fact, since $m_B=1$ is dummy message,  only $(m_{1},m_{2},\ldots,m_{B-1})$ need to be decoded.
For this, backwardly and sequentially for $g=B-1,B-2,\ldots, 1$, node $n+2$ finds the unique $m_g$ such that there exists some $l_{\tilde{\NN}, gB +1}^{gB+B }$ satisfying that for any block $j=gB +1,gB +2,\ldots,gB+B$,
\begin{align}
(&  \mathbf{X}_{0,j}(m_{f(j)}|m_{f(j-B)}),  \mathbf{X}_{1,j}(m_{f(j-B)})\nonumber   \\
& \{(\mathbf{X}_{i,j}(l_{i,j-1}) , \hat{\mathbf{Y}}_{i,j}(l_{i,j}|l_{i,j-1})): i \in \tilde{\NN} \}, \mathbf{Y}_{n+2,j}   )\in A_{\epsilon}(X_0,X_1,X_{\tilde{\NN} }, \hat{Y}_{\tilde{\NN} },Y_{n+2}). \label{E:chk''}
\end{align}
Note in  \dref{E:chk''}, for $j=gB +1,gB +2,\ldots,gB+B$, only $m_{f(j-B)}$, corresponding to $m_g$, needs decoding; and $m_{f(j)}$, corresponding to $m_{g+1}$, either is a dummy message (for $g=B-1$, i.e., $j=(B-1)B +1,(B-1)B +2,\ldots,B^2$), or has been decoded due to the backward property of decoding (for $g=B-2,\ldots, 1$). Thus, $X_0$ and $X_1$ are cooperatively transmitting the message $m_g$, and similarly as above, $m_g$ can be decoded if
\begin{align}
R<\min_{\SS \subseteq \tilde{\NN}}I(X_0,X_1,X_\SS;\hat Y_{\tilde{\NN} \setminus \SS},Y_{n+2}|X_{\tilde{\NN} \setminus \SS})-I(Y_{\SS};\hat Y_{\SS}|X_0,X_1,X_{\tilde{\NN}},Y_{n+2}, \hat Y_{\tilde{\NN} \setminus \SS}). \label{tobeimproved2}
\end{align}
Also, \dref{tobeimproved2} can be improved by considering only a subset $\TT_{n+2}$ for the decoding, leading to the following rate constraint:
\begin{align}
R<\max_{\TT_{n+2} \subseteq \tilde{\NN}}\min_{\SS \subseteq \TT_{n+2}}I(X_0,X_1,X_\SS;\hat Y_{\TT_{n+2} \setminus \SS},Y_{n+2}|X_{\TT_{n+2} \setminus \SS})-I(Y_{\SS};\hat Y_{\SS}|X_0,X_1,X_{\TT_{n+2}},Y_{n+2}, \hat Y_{\TT_{n+2} \setminus \SS}). \label{hasimproved2}
\end{align}

Combining  \dref{hasimproved1} and \dref{hasimproved2} and using the technique of time sharing, we have that the rate in \dref{E:simplified'} is achievable.

\subsection{Multiple D-F relay nodes ($M\geq 2$)}\label{subsection}

\emph{Codebook Generation:} Fix $p(x_0)p(x_{ \MM}|x_0)\prod_{i\in \NN \setminus \MM}p(x_i)p(\hat y_i|y_i,x_i)$. We randomly and independently generate a codebook for each block. The codebook generation for the C-F relay nodes is exactly the same as that in \ref{S:byb} and \ref{S:BbyBsingle}, and hence omitted. We only present the codebook generation for nodes $\pi(1:M+1)$. Still, denote $f(b):=\lceil \frac{b}{B}\rceil$, i.e., the smallest integer greater than or equal to $\frac{b}{B}$.
\begin{itemize}
  \item For each block $b\in [1: B^{M+1}]$, backwardly and sequentially for each relay node $\pi(k), k=M+1,M,\ldots,2$, randomly generate
$2^{TBR}$ conditionally independent sequences $$\mathbf{x}_{\pi(k),b}(m_{f(b-B^{k-1})}|m_{f(b-B^{k})},\ldots,m_{f(b-B^{M})}),$$ where $m_{f(b-B^{k-1})},m_{f(b-B^{k})},\ldots,m_{f(b-B^{M})}\in [1:2^{TBR}]$.
  \item For each block $b\in [1: B^{M+1}]$ and node $\pi(1)$, i.e., the source node 0, randomly generate $2^{TBR}$ conditionally independent sequences $$\mathbf{x}_{0,b}(m_{f(b)}|m_{f(b-B)},\ldots,m_{f(b-B^{M})}),$$ where $m_{f(b)},m_{f(b-B)},\ldots,m_{f(b-B^{M})}\in [1:2^{TBR}].$
\end{itemize}
The above, together with the codebook generation for the C-F relay nodes, defines the codebook for any block $b\in [1: B^{M+1}]$,
\begin{align*}
\mathcal{C}_b=\{&\mathbf{x}_{\pi(k),b}(m_{f(b-B^{k-1})}|m_{f(b-B^{k})},\ldots,m_{f(b-B^{M})}): m_{f(b-B^{k-1})},\ldots,m_{f(b-B^{M})}\in [1:2^{TBR}], k=M+1,M,\ldots,2; \\
&\mathbf{x}_{0,b}(m_{f(b)}|m_{f(b-B)},\ldots,m_{f(b-B^{M})}):m_{f(b)},m_{f(b-B)},\ldots,m_{f(b-B^{M})}\in [1:2^{TBR}];\\
 &\mathbf{x}_{i,b}(l_{i,b-1}) , \hat{\mathbf{y}}_{i,b}(l_{i,b}|l_{i,b-1}):  l_{i,b},l_{i,b-1} \in  [1:2^{T\hat R_i}], i\in \NN \setminus \MM       \}.
\end{align*}

\emph{Encoding:} Let the message vector to be sent be $$\mathbf{m}=(\underbrace{m_1,m_1,\ldots,m_1}_{B},\underbrace{m_2,m_2,\ldots,m_2}_{B},\ldots,\underbrace{m_{B^M},m_{B^M},\ldots,m_{B^M}}_{B}).$$
Let $m_{f(b)}=1$ be the dummy message for any
\begin{align}
b \in   \cup_{u=1}^{M} \cup_{v=1}^{B^{M-u}} [v(B-1)B^u +1:vB^{u+1}],\label{E:'dummy}
\end{align}
and for any $b\leq 0$. The actually achievable rate is $  (\frac{B-1}{B})^M R$ due to the dummy messages, which can still be made arbitrarily close to $R$ by letting $B\to \infty$ for any $M$.

The encoding process for the C-F relay nodes is still exactly the same as that in \ref{S:byb} and \ref{S:BbyBsingle}, and hence omitted. We only present the encoding process for nodes $\pi(1:M+1)$.
\begin{itemize}
  \item In block $b\in [1: B^{M+1}]$, the source node 0 transmits $\mathbf{x}_{0,b}(m_{f(b)}|m_{f(b-B)},\ldots,m_{f(b-B^{M})})$.
  \item At the end of block $vB^{k-1},  v\in [1:B^{M+1}/B^{k-1}]$, the relay node $\pi(k),k=2,\ldots,M+1$, has decoded messages $(m_1,m_2,\ldots,m_{vB^{k-2}})$ using backward decoding (see the decoding part). In the next $B^{k-1}$ blocks, i.e., in block  $b\in [vB^{k-1}+1:(v+1)B^{k-1}]$, the relay node $\pi(k),k=2,\ldots,M+1$, transmits $$\mathbf{x}_{\pi(k),b}(m_{f(b-B^{k-1})}|m_{f(b-B^{k})},\ldots,m_{f(b-B^{M})}),$$ where $(m_{f(b-B^{k-1})},m_{f(b-B^{k})},\ldots,m_{f(b-B^{M})})$ for any $b\in [vB^{k-1}+1:(v+1)B^{k-1}]$ have all been decoded by block $vB^{k-1}$.
\end{itemize}

\emph{Decoding:} At the end of every $B^{k-1}$ blocks, the node $\pi(k),k=2,\ldots,M+2$ decodes $B^{k-2}$ messages using $B$-Blocks-By-$B$-Blocks backward decoding. (Note every $B^{k-1}$ blocks carry $B^{k-2}$ messages.) Specifically, at the end of block $b =vB^{k-1},  v\in [1:B^{M+1}/B^{k-1}]$, the node $\pi(k),k=2,\ldots,M+2$, decodes messages $(m_{(v-1)B^{k-2}+1},\ldots,m_{vB^{k-2}})$.  In fact, $(m_{vB^{k-2}-B^{k-3}+1},\ldots, m_{vB^{k-2}})$ are dummy messages according to \dref{E:'dummy}, and only $(m_{(v-1)B^{k-2}+1},\ldots, m_{vB^{k-2}-B^{k-3}})$ need decoding.
For this, backwardly and sequentially for $g=vB^{k-2}-B^{k-3},vB^{k-2}-B^{k-3}-1,\ldots, (v-1)B^{k-2}+1$, node $\pi(k)$ finds the unique $m_g$ such that there exists some $l_{\NN\setminus \MM, (g-1)B+B^{k-2}+1}^{gB+B^{k-2}}$ satisfying that for any block $j=(g-1)B+B^{k-2}+1,(g-1)B+B^{k-2}+2,\ldots,gB+B^{k-2}$,
\begin{align}
(&  \mathbf{X}_{0,j}(m_{f(j)}|m_{f(j-B)},\ldots,m_{f(j-B^{M})}), \nonumber\\
&\{\mathbf{X}_{\pi(s),j}(m_{f(j-B^{s-1})}|m_{f(j-B^{s})},\ldots,m_{f(j-B^{M})}),s=2,\ldots,k-1,k,k+1,\ldots,M+1\}, \nonumber \\
& \{(\mathbf{X}_{i,j}(l_{i,j-1}) , \hat{\mathbf{Y}}_{i,j}(l_{i,j}|l_{i,j-1})): i \in \NN \setminus \MM \}, \mathbf{Y}_{\pi(k),j}   )\in A_{\epsilon}(X_0,X_\MM,X_{\NN \setminus \MM }, \hat{Y}_{\NN \setminus \MM },Y_{\pi(k)}), \label{E:chk'}
\end{align}
where $(m_{f(j)},m_{f(j-B)},\ldots,m_{f(j-B^{k-3})},m_{f(j-B^{k-2})},m_{f(j-B^{k-1})},\ldots, m_{f(j-B^{M})})$ are corresponding to
\begin{align}\label{messages}
(m_{g+B^{k-3}},m_{g+B^{k-3}-1},\ldots,m_{g+B^{k-3}-B^{k-4}},   m_{g},m_{g+B^{k-3}-B^{k-2}},\ldots, m_{g+B^{k-3}-B^{M-1}})
\end{align}
Among the messages in \dref{messages}, only $m_{g}$, corresponding to $m_{f(j-B^{k-2})}$, is the unknown message at node $\pi(k)$ that needs to be decoded. In fact, $$(m_{g+B^{k-3}-B^{k-2}},\ldots, m_{g+B^{k-3}-B^{M-1}}), \text{corresponding to } (m_{f(j-B^{k-1})},\ldots, m_{f(j-B^{M})}),$$
have been decoded by block $b-B^{k-1}$, while
$$(m_{g+B^{k-3}},m_{g+B^{k-3}-1},\ldots,m_{g+B^{k-3}-B^{k-4}}), \text{corresponding to }(m_{f(j)},m_{f(j-B)},\ldots,m_{f(j-B^{k-3})}),$$
either are dummy messages according to \dref{E:'dummy} (for $g=vB^{k-2}-B^{k-3}, \ldots,vB^{k-2}-2B^{k-3}+1$) or have been decoded due to the backward property of decoding (for $g=vB^{k-2}-2B^{k-3},vB^{k-2}-2B^{k-3}-1,\ldots,(v-1)B^{k-2}+1$). Therefore, in \dref{E:chk'},
$$\{\mathbf{X}_{\pi(s),j},s=k,k+1,\ldots,M+1\}$$ are known at node $\pi(k)$, while $$\{\mathbf{X}_{\pi(s),j},s=1,\ldots, k-1\}$$ are cooperatively transmitting the message $m_g$. Having noted this fact, from \cite[Thm 1]{KimElGamal} and its proof (see also \cite[Thm 2.4]{WuXieTIT}), we have that $m_g$ can be decoded if
\begin{align}
R<\min_{\SS \subseteq \NN\setminus \MM}I(X_{(1:k-1)},X_\SS;\hat Y_{(\NN\setminus \MM)\setminus \SS},Y_{\pi(k)}|X_{(\NN\setminus \MM)\setminus \SS},X_{(k:M+1)})-I(Y_{\SS};\hat Y_{\SS}|X_{(1:M+1)},X_{\NN\setminus \MM},Y_{\pi(k)}, \hat Y_{(\NN\setminus \MM)\setminus \SS}). \label{tobeimproved}
\end{align}
By considering only a subset $\TT_{k} \subseteq  \NN \setminus \MM$ for the decoding at node $\pi(k)$ while treating the inputs of other C-F relay nodes as purely noise, and using the technique of time sharing, \dref{tobeimproved} can be improved to
\begin{align}
R<\max_{\TT_k \subseteq \NN \setminus \MM} \min_{\SS \subseteq \TT_k} I(X_{\pi(1:k-1)},X_\SS;\hat Y_{\TT_k \setminus \SS},Y_{\pi(k)}|X_{\TT_k \setminus \SS},X_{\pi(k:M+1)},Q)-I(Y_{\SS};\hat Y_{\SS}|X_{\pi(1:M+1)},X_{\TT_k},Y_{\pi(k)}, \hat Y_{\TT_k \setminus \SS},Q), \end{align}
which proves Theorem \ref{main2}.

\section{Conclusion}\label{conclusion}
We have proposed a unified relay framework with both the D-F and C-F relay nodes for  multiple-relay channels. This framework employs nested blocks combined with backward decoding to allow for the full incorporation of the best known D-F and C-F relay strategies. The achievable rates obtained under such a framework turn out to combine both the best known D-F and C-F achievable rates and include them as special cases.

\end{document}